\documentclass[fleqn,usenatbib,useAMS]{mnras} 
\usepackage{graphicx}
\usepackage{color}
\usepackage{amsmath}
\usepackage{amssymb}
\usepackage{multicol}
\usepackage{bm}
\usepackage{pdflscape}

\usepackage[T1]{fontenc}
\usepackage{ae,aecompl}

\newcommand{\mearth}{M_\oplus}

\newcommand{\rsun}{R_{\rm \odot}}

\def\ms{\hbox{\,m\,s$^{-1}$}}         
\def\m2s2{\hbox{\,m$^{2}$\,s$^{-2}$}} 

\newcommand{\bluecomment}[1]{\textbf{\color{blue}#1}}

\title[The three super-Earths orbiting GJ9827]{Masses and radii for the three super-Earths orbiting GJ 9827, and implications for the composition of small exoplanets}

\author[Rice et al.]{K.~Rice$^{1,2}$\thanks{Email: wkmr@roe.ac.uk}, L.~Malavolta$^{3,4}$, A.~Mayo$^{5,6,7}$, A.~Mortier$^{8,9}$, L.A.~Buchhave$^{10}$, \newauthor L.~Affer$^{11}$, A.~Vanderburg$^{12,13,14}$, M.~Lopez-Morales$^{13}$, E. ~Poretti$^{15,16}$, \newauthor L. Zeng$^{17}$, A.C.~Cameron$^9$, M.~Damasso$^{18}$, A.~Coffinet$^{19}$, D. W.~Latham$^{13}$,  \newauthor A.S.~Bonomo$^{18}$, F.~Bouchy$^{19}$, D.~Charbonneau$^{13}$, X.~Dumusque$^{19}$, \newauthor P.~Figueira$^{20,21}$, A.F.~Martinez Fiorenzano$^{15}$, R.D.~Haywood$^{13,14}$, \newauthor J.~Asher Johnson$^{13}$, E.~Lopez$^{23,24}$, C.~Lovis$^{19}$,  M.~Mayor$^{19}$, G.~Micela$^{11}$, \newauthor E.~Molinari$^{15,22}$,  V.~Nascimbeni$^{4,3}$, C.~Nava$^{13}$, F.~Pepe$^{19}$, D. F.~Phillips$^{13}$, \newauthor G.~Piotto$^{4,3}$, D.~Sasselov$^{13}$, D.~S\'egransan$^{19}$, A.~Sozzetti$^{18}$, S.~Udry$^{19}$, \newauthor C.~Watson$^{25}$ \\ \\
$^1${SUPA, Institute for Astronomy, University of Edinburgh, Royal Observatory, Blackford Hill, Edinburgh, EH93HJ, UK} \\
$^2${Centre for Exoplanet Science, University of Edinburgh, Edinburgh, UK} \\
$^3$INAF - Osservatorio Astronomico di Padova, Vicolo dell'Osservatorio 5, 35122 Padova, Italy \\
$^4$Dipartimento di Fisica e Astronomia ``Galileo Galilei", Universita'di Padova, Vicolo dell'Osservatorio 3, 35122 Padova, Italy \\
$^5$Astronomy Department, University of California, Berkeley, CA 94720, USA\\
$^6$National Science Foundation Graduate Research Fellow \\
$^7$Fulbright Fellow \\
$^8$Astrophysics group, Cavendish Laboratory, University of Cambridge, J.J. Thomson Avenue, Cambridge CB3 0HE, UK \\
$^9$Centre for Exoplanet Science, SUPA, School of Physics and Astronomy, University of St Andrews, St Andrews KY169SS, UK \\
$^{10}$DTU Space, National Space Institute, Technical University of Denmark, Elektrovej 327, DK-2800 Lyngby, Denmark\\
$^{11}$INAF - Osservatorio Astronomico di Palermo, Piazza del Parlamento 1, I-90134 Palermo, Italy \\
$^{12}$Department of Astronomy, The University of Texas at Austin, 2515 Speedway, Stop C1400, Austin, TX 78712 \\
$^{13}$Harvard-Smithsonian Center for Astrophysics, 60 Garden Street, Cambridge, MA 02138, USA \\
$^{14}$NASA Sagan Fellow \\  
$^{15}$INAF - Fundaci\'on Galileo Galilei, Rambla Jos\'e Ana Fernandez P\'erez 7, 38712 Bre\~na Baja, Spain \\
$^{16}$INAF - Osservatorio Astronomico di Brera, Via E. Bianchi 46, 23807 Merate, Italy \\
$^{17}$Department of Earth and Planetary Sciences, Harvard University,
Cambridge, MA 02138, USA \\
$^{18}$INAF - Osservatorio Astrofisico di Torino, Via Osservatorio 20, I-10025 Pino Torinese, Italy \\
$^{19}$Observatoire de Gen\`eve, Universit\'e de Gen\`eve, 51 ch. des Maillettes, 1290 Sauverny, Switzerland \\
$^{20}$European Southern Observatory, Alonso de Cordova 3107, Vitacura, Santiago, Chile \\
$^{21}$Instituto de Astrof\'isica e Ci\^encias do Espa\c{c}o, Universidade do Porto, CAUP, Rua das Estrelas, 4150-762 Porto, Portugal\\
$^{22}$INAF - Osservatorio di Cagliari, via della Scienza 5, 09047 Selargius, CA, Italy\\
$^{23}$NASA Goddard Space Flight Center, 8800 Greenbelt Rd, Greenbelt, MD 20771, USA \\
$^{24}$GSFC Sellers Exoplanet Environments Collaboration, NASA GSFC, Greenbelt, MD 20771\\
$^{25}$Astrophysics Research Centre, School of Mathematics and Physics, Queen's University Belfast, Belfast BT7 1NN, UK \\
}

\date{Accepted XXX. Received YYY; in original form ZZZ}

\pubyear{2018}

\begin{document} 
\label{firstpage}
\pagerange{\pageref{firstpage}--\pageref{lastpage}}
\maketitle

\clearpage

 \begin{abstract}
 Super-Earths belong to a class of planet not found in the Solar System, but which appear common in the Galaxy. Given that some super-Earths are rocky, while others retain substantial atmospheres, their study can provide clues as to the formation of both rocky planets and gaseous planets, and - in particular - they can help to constrain the role of photo-evaporation in sculpting the exoplanet population. GJ 9827 is a system already known to host 3 super-Earths with orbital periods of 1.2, 3.6 and 6.2 days. Here we use new HARPS-N radial velocity measurements, together with previously published radial velocities, to better constrain the properties of the GJ 9827 planets. Our analysis can't place a strong constraint on the mass of GJ 9827 c, but does indicate that GJ 9827 b is rocky with a composition that is probably similar to that of the Earth, while GJ 9827 d almost certainly retains a volatile envelope.  Therefore, GJ 9827 hosts planets on either side of the radius gap that appears to divide super-Earths into pre-dominantly rocky ones that have radii below $\sim 1.5 R_\oplus$, and ones that still retain a substantial atmosphere and/or volatile components, and have radii above $\sim 2 R_\oplus$. That the less heavily irradiated of the 3 planets still retains an atmosphere, may indicate that photoevaporation has played a key role in the evolution of the planets in this system.
\end{abstract}

   
%

\vspace*{0.5cm}

\section{Introduction}
\label{intro}
One of the most exciting recent exoplanet results is the discovery that the most common type of exoplanet, with a period less than $\sim 100$ days, is one with a radius between that of the Earth ($1 R_\oplus$) and that of Neptune ($\sim 4 R_\oplus$) \citep{howard12,batalha13,fulton17,fulton18}.  Known as super-Earths, these appear to be common in the Galaxy, but are not found in our Solar System.  It also appears that the transition from being preferentially rocky/terrestrial to having a substantial gaseous atmosphere occurs within this size range \citep{rogers15}.  Recent studies \citep{fulton17,zeng17,vaneylen18} have suggested that there is in fact a gap in the radius distribution between 1.5 and $2 R_\oplus$, as predicted by \citet{owenwu13} and \citet{lopez13}.  Planets tend to have radii less than $\sim 1.5 R_\oplus$ and may be pre-dominantly rocky, or they sustain a substantial gaseous envelope and have radii above $2 R_\oplus$.

Super-Earths are, therefore, an important population as they may provide clues as to both the formation of gas giants and the formation of rocky, terrestrial planets.  In particular, they can help us to better understand the role that photo-evaporation plays in sculpting the exoplanet population.  It has been suggested that super-Earths probably formed with gas envelopes that make up at least a few percent of their mass \citep{rogers11,lopezfort14,wolfgang15}.  Those that are sufficiently strongly irradiated could then have lost their atmospheres via photo-evaporation \citep{lopez12,owenwu13,ehrenreich15}.  Those that have not been sufficiently strongly irradiated retain their atmospheres.  This could then explain the observed gap in the radius distribution \citep{owenwu17,fulton17,lopez18,vaneylen18}.  There may, however, be alternative explanations for this observed radius gap, such as late giant impacts \citep{inamdar15} or the atmosphere being stripped by the cooling rocky core \citep{ginzburg16,ginzburg18}.  This means that systems that have super-Earths on either side of this radius gap are particularly interesting. 

In this paper we present an analysis of one such system, the $K2$ target GJ 9827 (also known as K2-135, EPIC 246389858, or HIP 115752).  It is already known to host 3 super-Earths with radii between $1$ and $\sim 2 R_\oplus$, and with orbital periods of 1.21 days, 3.65 days, and 6.21 days \citep{niraula17, rodriguez18}.  \citet{rodriguez18} and \citet{niraula17} suggest that GJ 9827 b has a radius of $\sim 1.6 R_\oplus$, GJ 9827 c has a radius of $\sim 1.3 R_\oplus$, while GJ 9827 d has a radius of about $2 R_\oplus$. This means that these planets have radii that approximately bracket the radius gap detected by \citet{fulton17} which, as already suggested, makes this a particularly interesting system for studying the origin of this gap.  However, neither \citet{rodriguez18} nor \citet{niraula17} could independently estimate the planets masses and so used mass-radius relations \citep{weiss14,chen17}.

A recent radial velocity analysis \citep{teske18} has, however, presented mass estimates for the GJ 9827 planets. This analysis was unable to place strong constraints on the masses of GJ 9827 c and d, but suggests that GJ 9827 b has a mass of $\sim 8.2 \pm 1.53 M_\oplus$.  With a radius of $\sim 1.64 R_\oplus$ \citep{rodriguez18,niraula17}, this result would make GJ 9827 b one of the densest known super-Earths. This mass and radius would suggest that GJ 9827 b has an iron core that makes up a significant fraction of its mass, and could indicate that it has undergone a mantle-stripping collision with another body of a similar mass \citep{Marcus10}. A more recent analysis \citep{prieto-arranz18}, however, suggests that the mass of GJ 9827 b is not as high as suggested by \citet{teske18} and, in fact, may have a composition similar to that of the Earth.  This analysis also suggests that GJ 9827 c is also rocky, but that GJ 9827 d may retain a substantial, extended atmosphere.

Here we repeat the lightcurve analysis of GJ 9827 using the $K2$ data, which we present in Section \ref{sec:lightcurveanalysis}.  We also use the $K2$ lightcurve to constrain the stellar activity (Section \ref{sec:stellaractivity}). We then carry out a radial velocity analysis using the same radial velocity data as used by \citet{teske18}, \citet{niraula17} and \citet{prieto-arranz18}, but with an additional 41 new radial velocities from the HARPS-N spectrograph \citep{cosentino12,cosentino14}.   As we will discuss in Section \ref{sec:discussion}, we were able to constrain the masses of GJ 9827 b and d to better than ``10\%" and about ``20\%", but were not able to place a strong constraint on the mass of GJ 9827 c.  We also discuss what these results imply about the typical composition of planets below the radius gap \citep{fulton17}, a particular science goal of the HARPS-N Collaboration.   

\section{Radial Velocity Observations}
\label{sec:RVs}

\subsection{HARPS-N spectroscopy}
\label{sec:HNspect}
We collected a total of 43 radial velocity (RV) spectra of GJ 9827 with the HARPS-N spectrograph (${\rm R} = 115000$) installed on the 3.6-m Telescopio Nazionale Galileo (TNG) at the Observatorio de los Muchachos in La Palma, Spain \citep{cosentino12, cosentino14}. We observed GJ 9827 between August 2017 and December 2017 as part of the HARPS-N Collaboration's Guaranteed Time Observations (GTO) program. Our observational strategy consisted of taking one or two observations per night, separated by 2-3 hours, for several consecutive nights in order to properly sample the RV curve of all the transiting planets.

All the observations had an exposure time of 1800s. We eliminated one observation, taken on BJD=2458048.36, as it had an anomalously low signal-to-noise ratio (S/N) of less than 20, and another, taken on BJD=2457991.62, was rejected by the data reduction software because of abnormal flux correction. 

GJ 9827 has a V-band magnitude of V = 10.25, so, with the exception of the two observations that were eliminated, we obtained spectra with signal-to-noise ratios in the range S/N = 37 - 121 (average S/N = 70), at 550 nm in 30 minute exposures, resulting in an average RV precision of 1.9 \ms. 

\begin{figure}
\begin{center}
    \includegraphics[width=9.0cm]{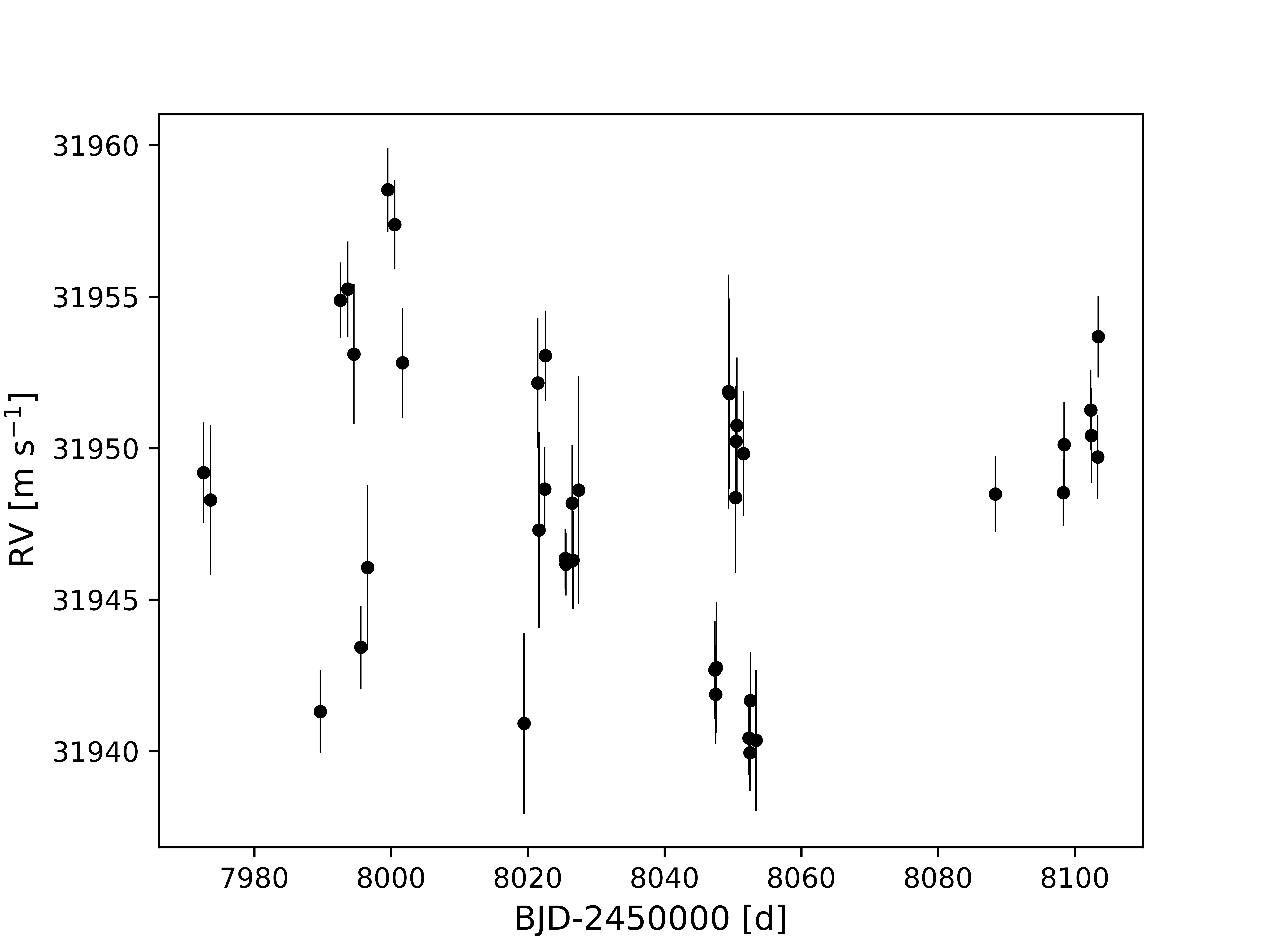}
    \caption{The HARPS-N radial velocities, first presented here, plotted against time.}
    \label{fig:HN_RVs}
\end{center}
\end{figure}

The spectra were reduced with version 3.8 of the HARPS-N Data Reduction Software (DRS), which includes corrections for color systematics introduced by variations in seeing \citep{cosentino14}.  The radial velocities were computed using a numerical weighted mask based on the synthetic spectrum of a K5 dwarf, following the methodology outlined in \citet{baranne96} and \citet{pepe02}. The HARPS-N data are presented in Table \ref{tab:harpsndata} and the radial velocities are shown in Figure \ref{fig:HN_RVs}. Table \ref{tab:harpsndata} also includes some stellar activity indicators.  Specifically, the full width at half maximum (FWHM) of the cross-correlation function (CCF), the line Bisector Inverse Slope (BIS), and an activity index derived from the Calcium H and K lines (${\rm S_{HK}}$).   

\subsection{Previously published HARPS-N and HARPS spectroscopy}
\label{sec:HSHN}
In our analysis we also use the HARPS-N and HARPS spectra first presented in \citet{prieto-arranz18}. 
This additional dataset includes 23 HARPS-N RV spectra, taken between July 2017 and December 2017, and 35 HARPS spectra taken between August 2017 and October 2017.
The HARPS instrument is installed on the 3.6m ESO telescope at La Silla and is very similar to the HARPS-N instrument, already discussed in Section \ref{sec:HNspect}. The HARPS and HARPS-N spectra were reduced with the same DRS as the HARPS-N spectra presented in Section \ref{sec:HNspect}. The HARPS-N RVs presented by \citet{prieto-arranz18} have an average precision of 1.6 \ms\ and signal-to-noise ratios in the range 33 - 95 (average S/N = 68), while the HARPS RVs have an average precision of of 1.4 \ms\ and signal-to-noise ratios in the range 47-100 (average S/N = 76). 

The \citet{prieto-arranz18} HARPS and HARPS-N data, including stellar activity indicators, can be found in their Tables 2 and 3, respectively. Their HARPS-N and HARPS RVs, together with the new HARPS-N RVs presented here, are shown in the top panel of Figure \ref{fig:HNHNHSPF_RVs}.  The filled circles show the new HARPS-N RVs from this study, the open circles show the HARPS-N RVs from \citet{prieto-arranz18}, and the open squares show their HARPS RVs. We also correct for the RV offset, but assume that all the HARPS-N data has the same offset\footnote{We verified that the two HARPS-N datasets were obtained using the same instrumental setup, and  analyzed with the same version of the  pipeline and using the same RV mask}, while allowing the HARPS and HARPS-N data to have different offsets.  This is discussed in more detail in Section \ref{sec:rvanalysis}.   

\subsection{Previously published Magellan/PFS and NOT/FIES spectroscopy}
\label{sec:PFSMagSpect}
In addition to the HARPS-N and HARPS spectra, we also include in our analysis the Magellan/PFS observations first presented by \citet{teske18}.  Thirty-six PFS observations were taken between January 2010 and August 2016, using the Planet Finder Spectrograph (PFS: \citealt{crane06}) on the {\it Magellan} II (Clay) Telescope.  The resolution was $\sim 80000$ and the exposure times were between 457 s and 900 s.  More details can be found in \citet{teske18}, and the resulting radial velocities are shown in their Table 1. 

Similarly, we also include the 7 high-resolution ($R \sim 67000$) spectra taken 
using the FIbre-fed \'Echelle Spectrograph (FIES: \citealt{telting14}) on the 2.6m Nordic Optical Telescope (NOT) of the Roque de los Muchachos Observatory (La Palma, Spain).  More details can be found in \citet{niraula17}, in which these observations were first presented, and the resulting radial velocities are shown in their Table 2.  

\begin{figure}
\begin{center}
    \includegraphics[width=9.5cm]{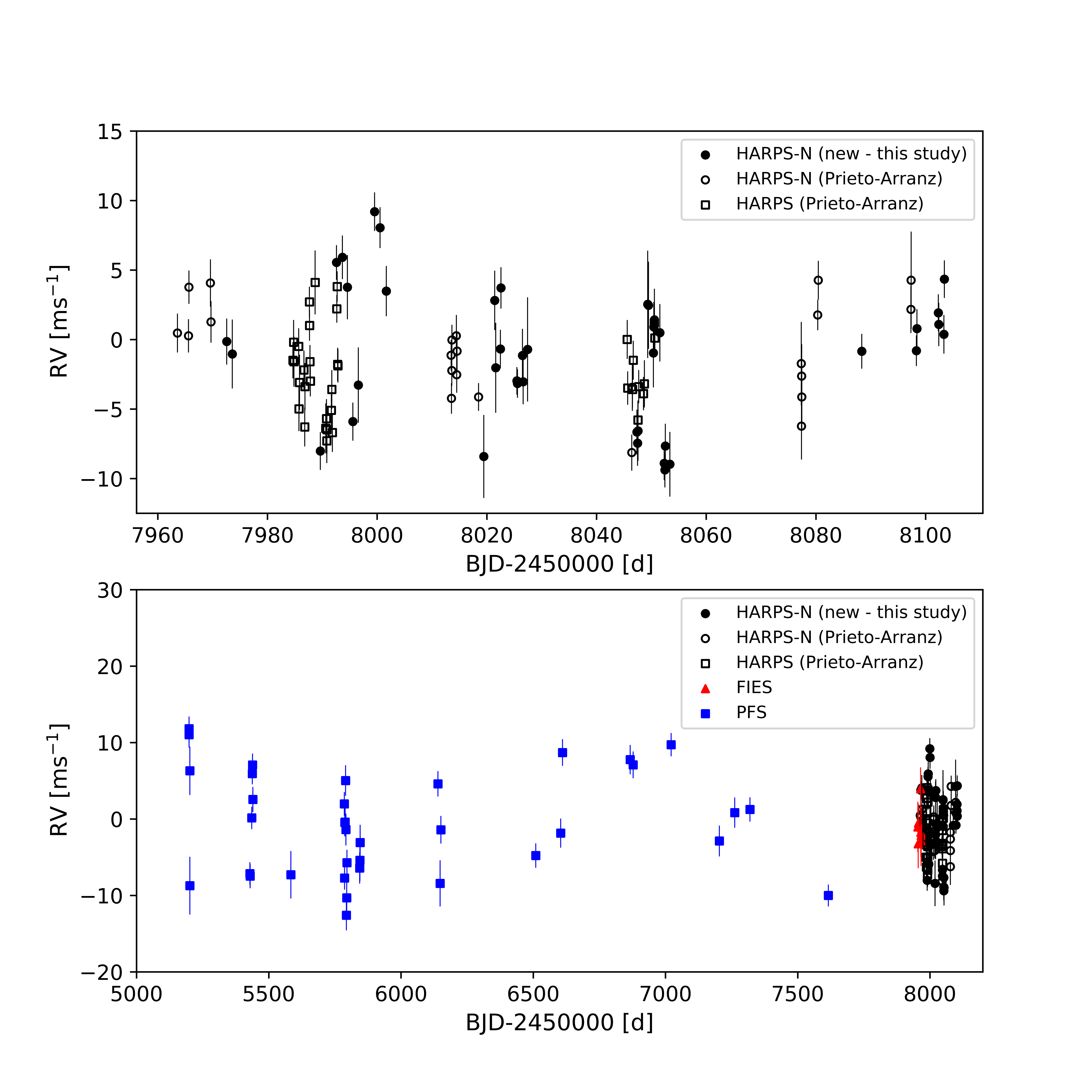}
    \caption{{\it Top panel:} The new HARPS-N RVs from this study (black filled circles) together with the HARPS-N and HARPS RVs presented by \citet{prieto-arranz18} (black open circles and black open squares respectively). {\it Bottom panel:} PFS (blue squares) and FIES (red triangles) RVs, together with all the HARPS-N and HARPS RVs (black filled circles, black open circles, and black open squares), all corrected for offsets between the datasets.}
    \label{fig:HNHNHSPF_RVs}
\end{center}
\end{figure}

The bottom panel of Figure \ref{fig:HNHNHSPF_RVs} shows the PFS (blue squares) and FIES (red triangles) radial velocities, together with the new HARPS-N RVs from this study (black filled circles) and the HARPS-N and HARPS RVs presented by \citet{prieto-arranz18} (black open circles and black open squares respectively).  These are all corrected for offsets between the datasets based on the best RV fit for the combined datasets discussed in Section \ref{sec:rvanalysis}.  

\section{Stellar parameters}
\label{sec:stellarparameters}
Taking advantage of the high S/N, high-resolution spectra obtained using HARPS-N, we re-determined the stellar parameters of GJ 9827 using the Stellar Parameter Classification pipeline (SPC; \citealt{buchhave14}). 

The high S/N needed to extract precise RVs means that these spectra are more than adequate for deriving stellar parameters.  Using the spectra obtained through the HARPS-N GTO program, we ran the SPC analysis on each individual spectrum, with a prior on the surface gravity from the YY isochrone models \citep{spada13}.  This SPC analysis yielded: $T_{\rm eff} = 4305 \pm 49$ K, $\log g = 4.72 \pm 0.10$ (cgs), [m/H] $= -0.50 \pm 0.08$ and $v \sin i < 2$ km s$^{-1}$.  The formal uncertainties also take into account the model uncertainties, which primarily stem from model systematics in the ATLAS Kurucz stellar models and degeneracies between the derived parameters when trying to compare observed spectra to model spectra \citep[see e.g.,][]{buchhave12,buchhave14}. 

To determine the mass, $M_\star$, and radius, $R_\star$, of GJ 9827, we used the  {\tt isochrones} Python package \citep{morton15}, which uses both the Mesa Isochrones and Stellar Tracks (MIST: \citealt{dotter16}) and the Dartmouth Stellar Evolution Database \citep{dotter08}. In addition to $T_{\rm eff}$, $\log g$ and [m/H], we included as priors the AAVSO Photometric All-Sky Survey $B$ and $V$ magnitudes \citep{henden15}, the 2MASS $J$ and $K$ magnitudes \citep{skrutskie06}, the WISE$2$ and $3$ magnitudes \citep{cutri14}, and the {\it Gaia} parallax from Data Release 2 \citep{gaia16,gaia18}.  We also repeat this analysis using the {\it Hipparcos} parallax \citep{vanleeuwen07}, to see how this impacts the resulting mass and radius estimates. We used both the MIST and Dartmouth model grids. Posterior sampling was performed using {\tt MultiNest} \citep{feroz08,feroz09,feroz13}.

In order to investigate the systematic errors on $M_*$ and $R_*$ introduced by the spectrally-derived stellar parameters
when dealing with late K and cooler dwarfs, we repeated the analysis using 
the stellar atmosphere parameters from \citet{niraula17}, \citet{teske18}, and \citet{rodriguez18}. \citet{niraula17} and \citet{teske18} used SpecMatch-Emp \citep{yee17}, the results of which are shown in their Table 3 and Table 2 respectively. The stellar parameters used by \citet{rodriguez18} are shown in their Table 1 and are taken from \citet{houdebine16}, who used principal component analysis.  

\begin{figure}
\begin{center}
    \includegraphics[width=9.25cm]{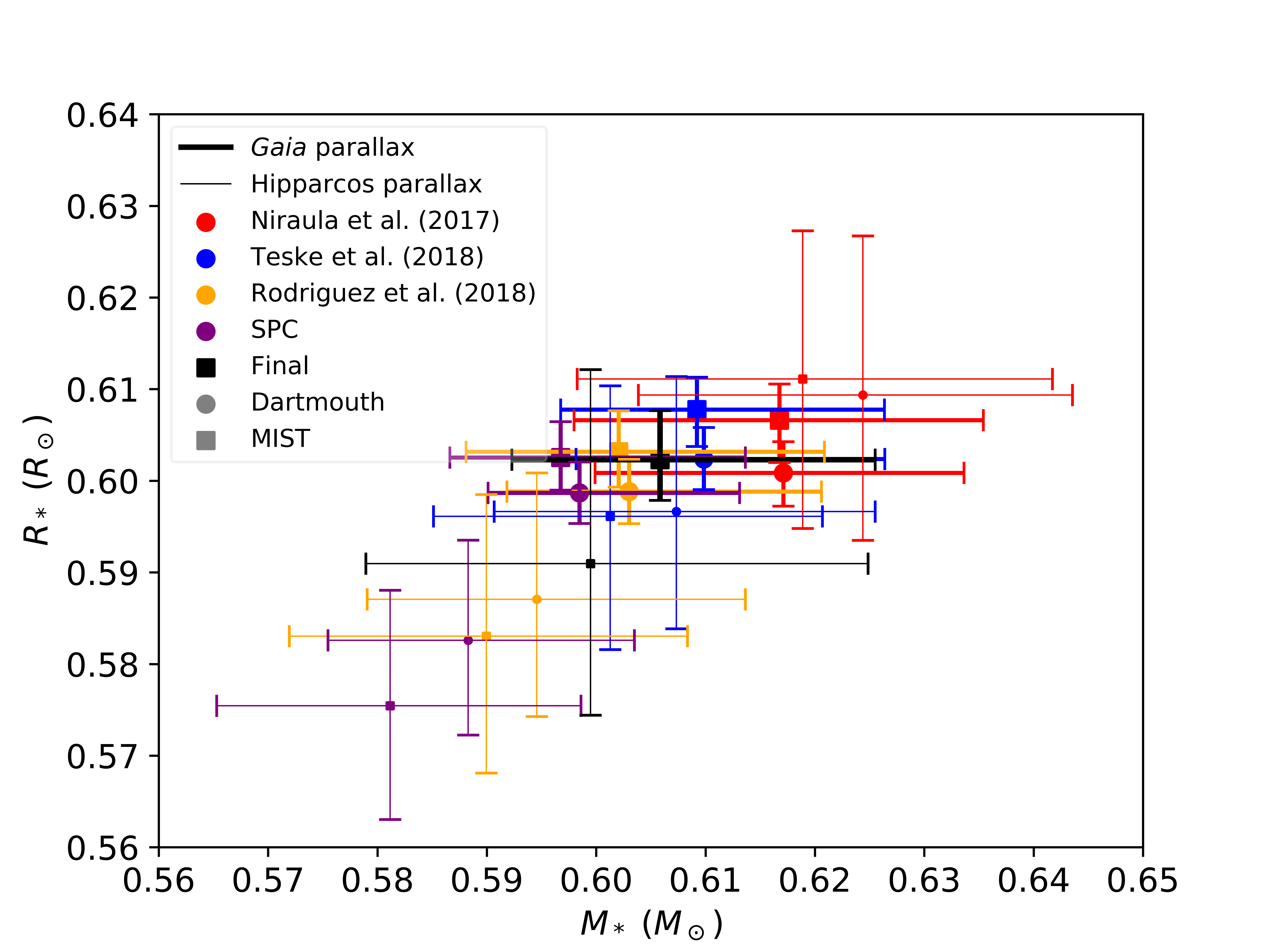}
    \caption{Mass ($M_*$) against radius ($R_*$) from our analysis using the {\tt isochrones} Python package. It includes results using our SPC analysis (purple), but also shows results using stellar parameters from \citet{niraula17} (red), \citet{teske18} (blue), and \citet{rodriguez18} (yellow).  We used both the MIST (squares) and Dartmouth (circles) model grids. We also use both the {\it Gaia} (thick lines) and {\it Hipparcos} (thin lines) parallaxes as priors. The black squares, with black error bars, show the mean $M_*$ and $R_*$.  It's clear that the {\it Gaia} parallax, which is more precise than the {\it Hipparcos} parallax, produces results that are more tightly constrained. For the rest of the analysis presented here, we will use the mean $M_*$ and $R_*$ determined using the {\it Gaia} parallax.}
    \label{fig:mass_radius}
\end{center}
\end{figure}

The results of our analysis are shown in Figure \ref{fig:mass_radius}. The thin lines show the results using the {\it Hipparcos} parallax as a prior, while the thick lines are the results obtained using the {\it Gaia} parallax as a prior. 

We then produce final estimates for each parameter by taking the median and  $15.865^{\rm th}/84.135^{\rm th}$ percentiles of the posterior samplings for all of the sets of stellar parameters, and for both the analysis using the {\it Gaia} parallax, and the analysis using the {\it Hipparcos} parallax.  The mass and radius obtained using the {\it Hipparcos} parallax as a prior are, $M_\star = 0.60^{+0.03}_{-0.02} M_\odot$ and $R_\star = 0.59^{+0.02}_{-0.02} R_\odot$, while using the {\it Gaia} parallax returns $M_\star = 0.606^{+0.020}_{-0.014} M_\odot$ and $R_\star = 0.602^{+0.005}_{-0.004} R_\odot$. It's clear that the more precise {\it Gaia} parallax produces results that are more tightly constrained than those obtained using the {\it Hipparcos} parallax.  Consequently, we use the results obtained with the \textit{Gaia} parallax for the rest of the analysis presented here.  Similarly, using the {\it Gaia} parallax, the {\tt isochrones} analysis returns $T_{\rm eff} = 4340^{+48}_{-53}$, ${\rm [m/H]} = -0.26 \pm 0.09$, $\log g = 4.66^{+0.015}_{-0.010}$ (cgs), and $A_v = 0.22 \pm 0.11$ for the effective temperature, metallicity, surface gravity, and interstellar reddening, respectively. The $T_{\rm eff}$ and $\log g$ results are consistent with the results from our SPC analysis, but the metallicity is discrepant at $2 \sigma$.  It is, however, consistent with some earlier metallicity estimates \citep{niraula17, teske18}. The $T_{\rm eff}$, $A_v$ and $R_*$ results are also consistent with results from {\it Gaia} Data Release 2 \citep{gaia16,gaia18}.  The {\tt isochrone} analysis also indicates that the star probably has an age of about 10 Gyr, with a lower limit (15.87th percentile) of 5Gyr.  

\begin{table}
\begin{center}
  \caption[]{GJ 9827 stellar parameters showing the magnitudes, results from our SPC analysis, and the {\it Hipparcos} and {\it Gaia} parallaxes used as input to the {\tt isochrones} Python package for estimating the mass, $M_*$, and radius $R_*$.  We don't show the stellar parameters from \citet{niraula17}, \citet{teske18}, and \citet{rodriguez18} that were also used in the {\tt isochrones} analysis (see text for details). Also shown are the metallicity, Effective temperature, $\log g$, and final mass and radius estimates obtained from the {\tt isochrones} analysis using the {\it Gaia} parallax as a prior.}
         \label{tab:stellarparameters}
   \begin{tabular}{lll}
            \hline\hline
            \noalign{\smallskip}
            Parameter     &  Description &  Value \\
            \noalign{\smallskip}
            \hline
            \noalign{\medskip}
            Other  & EPIC 246389858 & \\
            identifiers                  & HIP 115752  &  \\
                              & 2MASS J23270480-0117108 & \\
            \noalign{\smallskip}
            \hline\noalign{\smallskip}
            $B$ & APASS Johnson $B$ mag & $11.569 \pm 0.034$ \\
            $V$ & APASS Johnson $V$ mag & $10.250 \pm 0.138$ \\
            \noalign{\medskip}
            $J$ & 2MASS $J$ mag & $7.984 \pm 0.02$ \\
            $K$ & 2MASS $K$ mag & $7.193 \pm 0.02$ \\
            \noalign{\medskip}
            $WISE2$ & $WISE2$ mag & $7.155 \pm 0.02$ \\
            $WISE3$ & $WISE3$ mag & $7.114 \pm 0.017$ \\
            \noalign{\medskip}
            $v \sin i$ & Rotational Velocity (SPC) & $< 2$ km s$^{-1}$ \\
            ${\rm [m/H]}$ & Metallicity (SPC) & $-0.5 \pm 0.08$ \\
            $T_{\rm eff}$  &  Effective Temperature (SPC) & $4305 \pm 49$ K \\
            $\log g$ & Surface Gravity (SPC) & $4.72 \pm 0.1$ (cgs) \\
            \noalign{\medskip}
            $\pi_{Hip}$ & {\it Hipparcos} Parallax (mas) & $32.98 \pm 1.76$ \\
            $\pi_{GAIA}$ & {\it Gaia} Parallax (mas) & $33.68 \pm 0.06$ \\
            \noalign{\medskip}
            ${\rm [m/H]}$ & Metallicity ({\tt isochrones}) & $-0.26 \pm 0.09$ \\
            $T_{\rm eff}$ & Effective temperature & $4340^{+40}_{-53}$ K\\
             & ({\tt isochrones}) & \\
            $\log g$ & surface gravity ({\tt isochrones}) & $4.66^{+0.015}_{-0.010}$ (cgs) \\
            \noalign{\medskip}
            $M_*$ & Mass ({\tt isochrones}) & $0.606^{+0.020}_{-0.014}$ M$_\odot$ \\
            \noalign{\smallskip}
            $R_*$ & Radius ({\tt isochrones}) & $0.602^{+0.005}_{-0.004}$ R$_\odot$ \\
            \noalign{\smallskip}
            \hline
    \end{tabular}    
\end{center}
\end{table}

\subsection{Stellar kinematics} \label{sec:kinematic}
Stars presently near the Sun may come from a wide range of Galactic locations.
Therefore, stellar space velocity, as a clue to the origin of a star in the
Galaxy, is very important. The accurate {\it Gaia} parallax (see Table \ref{tab:stellarparameters}), combined with the proper motions and the stellar radial velocity, make
it possible to derive reliable space velocities for GJ\,9827. The calculation of the space
velocity with respect to the Sun is based on the procedure presented by \citet{johnson87}, corrected for the effect of differential galactic rotation \citep{scheffler87}, by adopting a solar
Galactocentric distance of 8.5 kpc and a circular velocity of 220 km
s$^{-1}$. The correction of space velocity to the Local Standard of Rest is
based on a solar motion\footnote{In the present work, $\vec{U}$ is defined to be positive in the
direction of the Galactic center.}, $(U, V, W)_{\sun}=(10.0, 5.2, 7.2)$ km s$^{-1}$, as derived from
{\it Hipparcos} data by \citet{dehnen88}.
The peculiar space velocity $S$, given by $S=(U^2+V^2+W^2)^{1/2}$, is
quoted with all kinematic data in Table \ref{tab:kinematics} (with the exception of the {\it Gaia} parallax which is included in Table \ref{tab:stellarparameters}). GJ 9827,
shows kinematic properties typical of the thin disk population. We have calculated the probabilities that the star belongs to a specific
population, thick (TD), thin disk (D) or stellar halo (H), following the method used
by \citet{bensby04}. On account of these probabilities, we find for
GJ 9827 a thick-disk to thin-disk probability ratio of $TD/D=0.05$, implying that the star is clearly identified as a
thin-disk object (typical threshold for assignment to thin disk being TD/D less than 0.1).

\begin{table}
\caption{Kinematic data.}
\label{tab:kinematics}
\begin{center}
\begin{tabular}{cc}
\hline\hline     
Parameter & GJ\,9827\\\hline
$\mu_{\alpha}$[mas/yr]$^{(1)}$& 376.02$\pm$0.06\\
$\mu_{\beta}$[mas/yr]$^{(1)}$& 216.07$\pm$0.07\\
$U_{LSR}$ [km s$^{-1}$]$^{(2)}$ & -49.4$\pm$0.4 \\
$V_{LSR}$ [km s$^{-1}$]$^{(2)}$ &  22.9$\pm$0.9 \\  
$W_{LSR}$ [km s$^{-1}$]$^{(2)}$ & -18.6$\pm$1.1 \\   
S [km s$^{-1}$]$^{(2)}$ &  57.5$\pm$0.6\\\hline\hline                 
\end{tabular}
\begin{flushleft}
References. $^{(1)}$ Gaia Collaboration et al. 2016, 2018; $^{(2)}$ This work (see text).
\end{flushleft}
\end{center}
\end{table}

\section{$K2$ Photometry and light curve analysis}\label{sec:lightcurveanalysis}
After the failure of the second of its four reaction wheels, the $Kepler$ spacecraft was re-purposed for an extended $K2$ mission to obtain high-precision photometry on a set fields near the ecliptic. GJ 9827 was observed from UT 2016 December 16 until UT 2017 March 04, as part of K2 campaign 12.  

Our data reduction and analysis techniques is very similar to that described in Sections 2.2 and 4.1 of \citet{mayo18}. We also provide a summary of our methods here. We first applied the method developed by \citet{vanderjohns14} and \citet{vanderburg16b} in order to remove the roll systematics introduced by the periodic thruster firing of the {\em Kepler Space Telescope}. Next, we removed low-frequency variations from the light curve via a basis spline. Then we used the BATMAN transit model \citep{kreidberg15} to simultaneously fit the transits of all three planets, assuming non-interaction and circular orbits. The latter assumption seems reasonable, given that the system is old enough for tidal circularisation to have occured \citep{barnes17}, and that systems similar to GJ 9827 do tend to have low eccentricities \citep{vaneylen15}.  Additionally, as will be discussed in Section \ref{sec:rvanalysis}, the RV analysis is also consistent with the planets having circular orbits.

The model included four global parameters: baseline flux level, a noise parameter, and two quadratic limb darkening coefficients parameterized according to \citet{kipping13}. Unlike \citet{mayo18} we also impose a stellar density prior of $3.92 \pm 0.014$ g cm$^{-3}$, determined using the stellar mass and radius determined in Section \ref{sec:stellarparameters}.  When imposing this prior, we also assume that the three planets each have circular orbits.  

Additionally, each planet had five parameters: the \bluecomment{initial} epoch (i.e. time of first transit), the period, the inclination, the ratio of planetary to stellar radius ($R_p/R_*$), and the semi-major axis normalized to the stellar radius ($a/R_*$). All parameters were given a uniform prior except for each planet's $R_p/R_*$, for which we assumed a log-uniform prior. We estimated these model transit parameters using {\tt emcee} \citep{foreman13}, a Python package which performs Markov chain Monte Carlo (MCMC) simulations with an affine-invariant ensemble sampler \citep{goodman10}. Using 38 walkers (i.e. twice the number of model parameters), we ran the MCMC process until convergence, which we defined as the point at which the scale-reduction factor \citep{gelman92} dropped below 1.1 for every parameter.

\begin{figure}
\begin{center}
    \includegraphics[width=8.5cm]{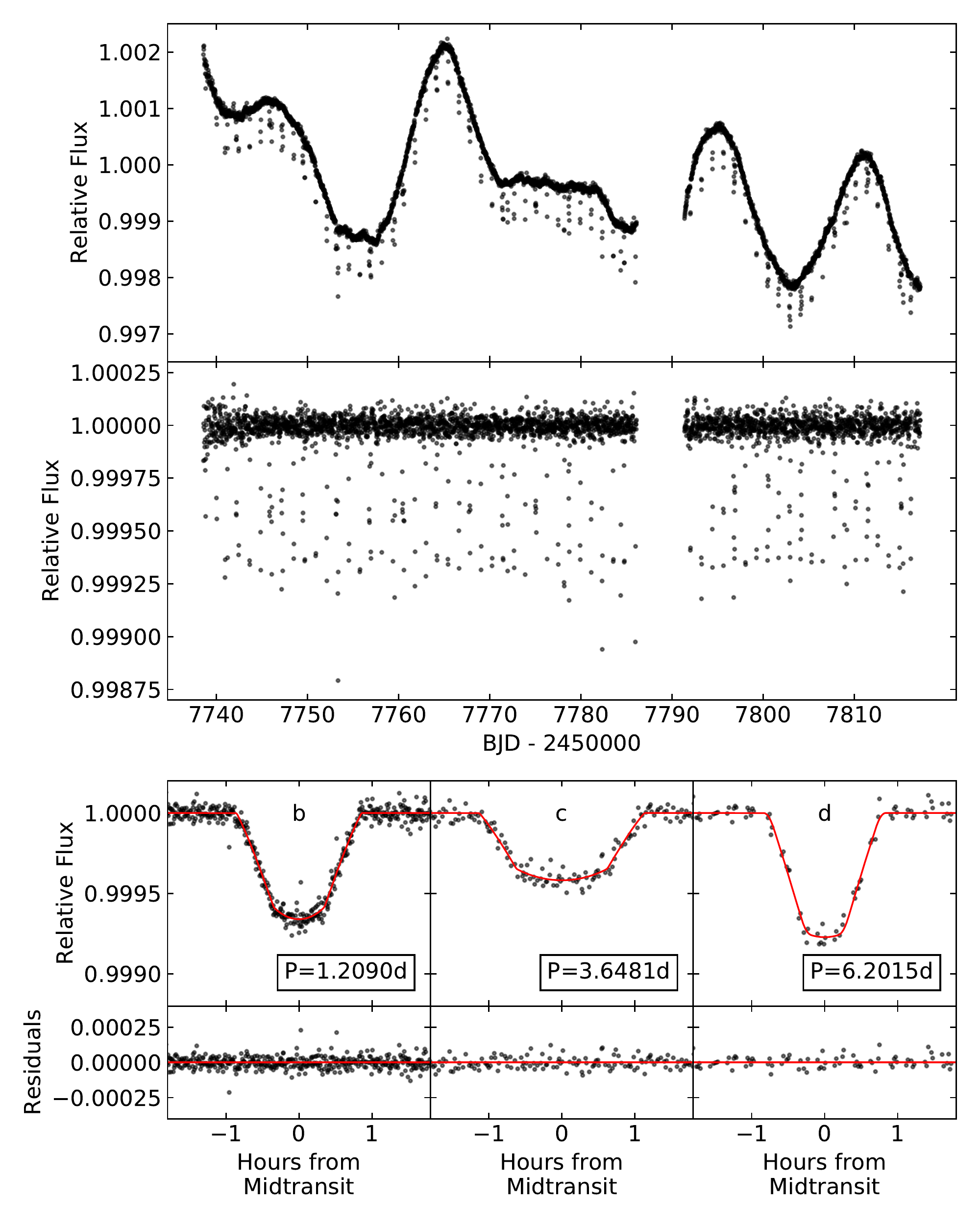}
    \caption{{\it Top panel:} $K2$ lightcurve after removing the roll systematics introduced by the periodic thruster fires of the {\em Kepler Space Telescope}, but without the removal of the low-frequency variations. {\it Middle panel:} $K2$ lightcurve after also removing the low-frequency variations. {\it Bottom panel:} Phase-folded lightcurves for planets b, c, and d.  The model is shown in red and the residuals are shown in the lower parts of the panel.}
    \label{fig:lightcurveanalysis}
\end{center}
\end{figure}

The systematics corrected, normalised, and phase folded lightcurves are shown in Figure \ref{fig:lightcurveanalysis}.  The results of our lightcurve analysis are shown in Table \ref{tab:lightcurveanalysis}.  For completeness, the baseline flux level is $1.000 \pm 0.000002$, the noise parameter is $\log(jitter) = -10.11 \pm 0.002$, and the quadratic limb darkening parameters are $q_1 = 0.3999^{+0.2403}_{-0.1642}$ and $q_2 = 0.4372^{+0.3004}_{-0.2173}$. Our results agree well with those in \citet{rodriguez18} and \citet{niraula17}, and suggest that GJ 9827 b and d, with radii of $R_{p,b} = 1.577^{+0.027}_{-0.031}$ and $R_{p,d} = 2.022^{+0.046}_{-0.043}$, roughly lie on either side of the radius gap detected by \citet{fulton17}. The derived quantities in Table \ref{tab:lightcurveanalysis} ($R_p$ and $a$) were determined by sampling the posterior distributions of the dependent quantities, and presenting the median of the resulting distribution with the uncertainties being the difference between this median value and the 16$^{\rm th}$ and 84$^{\rm th}$ percentile values.   

\begin{table*}
\begin{center}
  \caption[]{Planetary parameters from the light curve analysis.}
         \label{tab:lightcurveanalysis}
   \begin{tabular}{lllcc}
            \hline\hline
            \noalign{\smallskip}
            Parameter     &  Description &  GJ 9827 b & GJ 9827 c & GJ 9827 d \\
            \noalign{\smallskip}
            \hline
            \noalign{\medskip}
            $P$ & Period (days) & $1.20898190^{+0.00000693}_{-0.00000714}$ & $3.6480957^{+0.0000633}_{-0.0000621}$ & $6.2014698^{+0.0000626}_{-0.0000611}$ \\
            \noalign{\smallskip}
            $R_p/R_*$ & Radius of the planet in stellar radii & $0.02396^{+0.00037}_{-0.00044}$ & $0.01887^{+0.00034}_{-0.00037}$ & $0.03073^{+0.00065}_{-0.00060}$ \\
            \noalign{\smallskip}
            $R_p$ & Radius of the planet (R$_\oplus$)$^a$ &  $1.577^{+0.027}_{-0.031}$ & 
            $1.241^{+0.024}_{-0.026}$ & $2.022^{+0.046}_{-0.043}$ \\ 
            \noalign{\smallskip}
            $T_C$ & Time of Transit (BJD-2454833) & $2905.82586^{+0.00026}_{-0.00026}$ & $2909.19930^{+0.00072}_{-0.00073}$ & $2907.96115^{+0.00044}_{-0.00045}$ \\
            \noalign{\smallskip}
            $T_{\rm 14}$ & Transit duration (days) & $0.05270^{+0.00093}_{-0.00083}$ & $0.07604^{+0.00154}_{-0.00154}$ & $0.05095^{+0.00147}_{-0.00122}$ \\
             \noalign{\smallskip}
            $b$ & Impact parameter & $0.4602^{+0.0352}_{-0.0443}$ & $0.4428^{+0.0415}_{-0.0483}$ & $0.8927^{+0.0071}_{-0.0090}$ \\
            \noalign{\smallskip} $i$ & Inclination & $86.07^{+0.41}_{-0.34}$ & $88.19^{+0.21}_{-0.18}$ & $87.443^{+0.045}_{-0.045}$ \\
            \noalign{\smallskip}
            $a/R_*$ & Semimajor axis in stellar radii & $6.719^{+0.080}_{-0.086}$ & $14.035^{+0.172}_{-0.171}$ & $20.003^{+0.230}_{-0.254}$ \\
            \noalign{\smallskip}
            $a$ & Semimajor axis (au)$^b$ & $0.01880^{+0.00020}_{-0.00014}$ & $0.03925^{+0.00042}_{-0.00029}$ & $0.05591^{+0.00059}_{-0.00041}$ \\
             \noalign{\medskip}
            \hline
    \end{tabular}    
\begin{flushleft}
{\bf Notes.}$^{a}$ Radii are derived using our estimate for the stellar radius, $R_*=0.602^{+0.005}_{-0.004} \rsun$, and the ratios $R_{\rm planet}/R_{\rm star}$ determined here. $^b$ Semimajor axes are determined  assuming that  $M_{\rm s}+m_{\rm p} \cong M_{\rm s}$ and using $a \cong [(M_{\rm s}\cdot G)^{\frac{1}{3}}\cdot P_{\rm p}^{\frac{2}{3}}]/(2\pi)^{\frac{2}{3}} $, where $G$ is the gravitational constant.
\end{flushleft}
\end{center}
\end{table*}

\section{Stellar activity}
\label{sec:stellaractivity}
Characterizing the activity level of the host star, and eventually modelling the activity contribution to the RV, is mandatory for accurate mass determination of small planets, even when the star is just moderately active (e.g.,  \citealt{haywood2018}). The $K2$ light curve shows a strong modulation with peak-to-peak amplitude of $\simeq 0.003$ mag, suggesting a non-negligible level of activity for this star.

Previous analyses have estimated GJ 9827's rotation period, but the results are not consistent. \citet{niraula17} suggests a rotation period of $\sim 17$ days, while \citet{rodriguez18} and \citet{teske18} suggest a rotation period of 31 days. 

Correcting for activity induced signals in the radial velocity data requires an accurate estimate of the star's rotational period. We use the combined HARPS and HARPS-N dataset to carry out a periodogram analysis of the BIS and the FWHM of the CCF, as computed by the DRS (see Section \ref{sec:HNspect}), and the activity index derived from the Calcium H and K lines (${\rm S_{HK}}$, see Table \ref{tab:harpsndata} and Tables 2 and 3 in \citealt{prieto-arranz18}). Specifically, we used the Bayesian formalism for the generalised Lomb-Scargle periodogram first presented by \citet{mortier15}. The spectral window of the HARPS and HARPS-N data shows a peak at
$\sim$27~days due to the Moon's sidereal month. This hampers our ability 
to best exploit our data to derive a reliable measure
of the stellar rotational period. When analysing the activity indices, 
a significant signal is, however, found in the ${\rm S_{HK}}$ data at  
$\sim$34~days with another peak at around $\sim 15$ days. We also consider correlations between
the activity indices and the RVs. The Spearman's rank
correlation coefficients are all below 0.3.

Therefore, we also carry out a frequency analysis of the combined HARPS and HARPS-N
RV data using the Iterative Sine-Wave Fitting (ISWF) method
\citep{vanicek71}. The power spectra 
shows clear peaks at $f_b$=0.827~d$^{-1}$ (corresponding to
the orbital period of GJ 9827 b, $P_b$=1.209~days) and 
at $f_d$ = 0.161~d$^{-1}$ (corresponding 
to the orbital period of GJ 9827 d, $P_d = 6.21$ days). The low-amplitude signal due to GJ 9827 c can be 
seen in the power
spectrum, but it does not stand out above the noise.
This frequency analysis also shows peaks at $f$=0.0325~d$^{-1}$,
2$f$, 3$f$. The frequency $f$ corresponds to a period of 30.8 days 
and is clearly related to the stellar rotation period.  This would seem to indicate that the $\sim 15$ day signal
seen in the ${\rm S_{HK}}$ data is probably the first harmonic of the stellar rotation period.

To better quantify the stellar activity, we carry out an analysis using the $K2$ light curve (see top panel of Figure \ref{fig:lightcurveanalysis}) but after removing the points affected by transits.  We initially determined the auto correlation of the $K2$ light curve data, computed as described in \citet{mcquillan13}\footnote{As implemented in \url{https://github.com/bmorris3/interp-acf}.}. This converges to a rotational period of 29 days, which is closer to the $31$ days presented in \citet{rodriguez18} and \citet{teske18}, than to the $\sim 17$ days suggested by \citet{niraula17}. 

It has, however, been suggested \citep{angus18} that a Gaussian process (GP) with a quasi-periodic covariance kernel function is a more reliable method to determine the rotational period of active stars.  We therefore performed an additional analysis using {\tt PyORBIT}\footnote{Version 5, available at \url{https://github.com/LucaMalavolta/PyORBIT}.} \citep{malavolta16}, a package for modelling planetary and activity signals. This implements the GP quasi-periodic kernel through the {\tt george} package \citep{ambikasaran15}. For the hyper-parameters we follow the mathematical definition introduced by \citet{grunblatt15}. Hyper-parameters optimization has been performed using the differential evolution code {\tt pyDE}\footnote{Available at \url{https://github.com/hpparvi/PyDE}}, which provided the starting values for the affine-invariant ensemble sampler {\tt emcee} \citep{foreman13}. We followed the same methodology as described in \cite{malavolta18}. 

Since the GP regression typically scales with the third power of the number of data points, we binned the $K2$ light curve every 5 points, while ensuring that this did not alter the overall shape and did not change the auto correlation result. Since there is a data gap between BJD - 2450000 = 7786 and BJD - 2450000 = 7791, we also allow for different offsets and jitters for the two data segments.  The GP analysis then suggested a rotational period of $P_{\rm rot} = 28.72^{+0.18}_{-0.22}$ days, a decay timescale of the active regions of $\lambda = 33.17^{+5.90}_{-6.26}$ days, and a coherence scale of $w = 0.146 \pm 0.006$.  We also find a covariance amplitude in the $K2$ light curve data of $h_{K2} = 0.00081^{+0.00013}_{-0.00010}$ mag. These values are also presented in Table \ref{tab:stellaractivity}. 

The GP regression therefore produces a result that is consistent with that from the auto correlation of the $K2$ light curve data and with that presented in \citet{rodriguez18} and \citet{teske18}.  The isochrone analysis also suggests that this star has an age of $\sim 10$ Gyrs, with a lower limit of 5 Gyrs.  The stellar kinematics, reported in Section \ref{sec:kinematic}, indicates that GJ 9827 belongs to the galactic thin disk, but the low metallicity (${\rm [m/H]} = -0.26 \pm 0.09$) is consistent with this being an older member of that population.  A rotation period of $\sim 30$ days is consistent with what would be expected for a star of this age \citep{reiners12}. This would all seem to indicate that the rotation period of GJ 9827 is more likely $\sim 30$ days than the $\sim 17$ days suggested by \citet{niraula17}.  Therefore we will use the results of the GP regression to correct for the stellar activity induced signals in the radial velocity data.  

\begin{table}
\begin{center}
  \caption[]{Stellar activity indicators from the GP model using the $K2$ lightcurve only.}
         \label{tab:stellaractivity}
   \begin{tabular}{lll}
            \hline\hline
            \noalign{\smallskip}
            Parameter     &  Description &  Value \\
            \noalign{\smallskip}
            \hline
            \noalign{\medskip}
            $\sigma_{1, \rm jit, K2}$$^a$ [mag] & Jitter & $0.000006^{+0.000003}_{-0.000003}$ \\
            \noalign{\smallskip}
            $\sigma_{2, \rm jit, K2}$$^a$ [mag] & Jitter & $0.000003^{+0.000003}_{-0.000002}$ \\
            \noalign{\smallskip}
            $\gamma_{1, \rm K2}$$^a$ [mag] & Offset & $1.000237^{+0.000263}_{-0.000260}$ \\
            \noalign{\smallskip}
            $\gamma_{2, \rm K2}$$^a$ [mag] & Offset & $0.999427^{+0.000273}_{-0.000268}$ \\
            \noalign{\medskip}
            $P_{\rm rot}$ [days] & Rotational period & $28.72^{+0.18}_{-0.22}$ \\
            \noalign{\smallskip}
            $\lambda$ [days] & Active region & $33.17^{+5.90}_{-6.26}$  \\
              & decay timescale & \\
            \noalign{\smallskip}
            $w$ [mag] & Coherence Scale & $0.146^{+0.006}_{-0.006}$ \\
            \noalign{\smallskip}
            $h_{K2}$ & Covariance amplitude & $0.00081^{+0.00013}_{-0.00010}$ \\
             \noalign{\medskip}
            \hline
    \end{tabular}   
\begin{flushleft}
$^a$ The terms $\sigma_{1, \rm jit, K2}$ and $\gamma_{1, \rm K2}$ are for the K2 data segment ending at BJD-2450000=7786.075, while $\sigma_{2, \rm K2}$ and $h_{2, \rm K2}$ are for the K2 data segment starting at BJD-2450000=7791.377.
\end{flushleft}
\end{center}
\end{table}

\section{RV Analysis}
\label{sec:rvanalysis}
The $K2$ light curve analysis, the ISWF analysis of the HARPS and HARPS-N RVs, and the ${\rm S_{HK}}$ index clearly suggest that the stellar activity of GJ 9827 may have non negligible effects on the RVs. The approach that we've taken is to assume that the light curve variations and activity signals in the RVs can be described by a GP with the same kernel and with common hyper-parameters, except for the covariance amplitude, $h$, which is specific for each dataset. This approach has been quite successful in confirming and improving mass determination of rocky planets (e.g., \citealt{haywood14}, \citealt{grunblatt15}), and it has delivered consistent results with respect to alternative approaches for stellar activity modelling (e.g. \citealt{malavolta18}).  In the context of the GP analyis, we take the combined HARPS and HARPS-N RVs to be a single dataset, with the PFS RVs and FIES RVs making up two other datasets.  We do, however, allow for an offset between the HARPS and HARPS-N RVs and for independent jitter terms.

We carry out the RV analysis using the {\tt PyORBIT} code and, as in Section \ref{sec:stellaractivity}, assume that the quasi-periodic kernel is the best choice to model RV variations. When modelling activity signals in RVs with the help of Gaussian processes in a Bayesian framework, 
imposing priors obtained from the $K2$ lightcurve on the hyper-parameters of the GP produces statistically indistinguishable results when compared to modelling the RVs and the lightcurve simultaneously \citep[e.g., ][]{malavolta18}. 

Hence, rather than modelling the $K2$ light curve and the RVs simultaneously, we use the results from Section \ref{sec:stellaractivity} to set priors on the hyper-parameters, with the exception of the amplitude of the covariance $h$. Since the RV intensity of stellar activity depends on the wavelength range of the instrument and the RV extraction technique \citep[e.g., ][]{zechmeister18}, for each dataset (combined HARPS and HARPS-N RVs, PFS RVs, and FIES RVs) we used an independent covariance amplitude $h$. For each dataset we also include a jitter term, to compensate for uncorrelated noise not included in the error estimate, and a RV offset.  As mentioned above, although we treat the combined HARPS and HARPS-N RVs as a single dataset, we do allow for different offsets and jitter values for the HARPS and HARPS-N RVs. We use uniform priors for both the jitter and the RV offset. We ran two main analyses, one in which the results from Section \ref{sec:stellaractivity} exactly define the Gaussian priors on the hyper-parameters, and one in which they guide our choice of priors, but do not precisely define them.  Specifically, in the second analysis we use Gaussian priors on $P_{\rm rot}$, $\lambda$, and $w$, with $P_{\rm rot}=35 \pm 10$ days, $\lambda = 36 \pm 15$ days, and $w = 0.15 \pm 0.005$. 

In the second analysis, we set the $P_{\rm rot}$ prior to the value we would have used if only spectroscopic activity indexes had been used to estimate the stellar rotation period ($\sim 35$ days from the ${\rm S_{HK}}$ index, see Section \ref{sec:stellaractivity}), but we make the range wide enough to also incorporate the results from the analysis using the $K2$ lightcurve and to account for the photometry and RVs not being simultaneous in time.

In our model we also assume that the orbits of all 3 planets are circular (eccentricity $e = 0$). In multi-planet systems of close-in planets, the eccentricity evolution depends on both tidal interactions and on eccentricity pumping from planet-planet interactions \citep{bolmont13}. However, given the age of the system ($> 5$ Gyr) there has probably been sufficient time for the orbits of these close-in planets to have been tidally circularised \citep{barnes17}, and there are indications that systems like GJ 9827 do tend to have low eccentricities \citep{vaneylen15}. We also impose a log-uniform prior on the time of transit centre and a uniform prior on the orbital periods of the 3 planets, taken from the results of the analysis discussed in Section \ref{sec:lightcurveanalysis}  (see Table \ref{tab:lightcurveanalysis}). 

\begin{figure*}
\begin{center}
    \includegraphics[width=17.0cm]{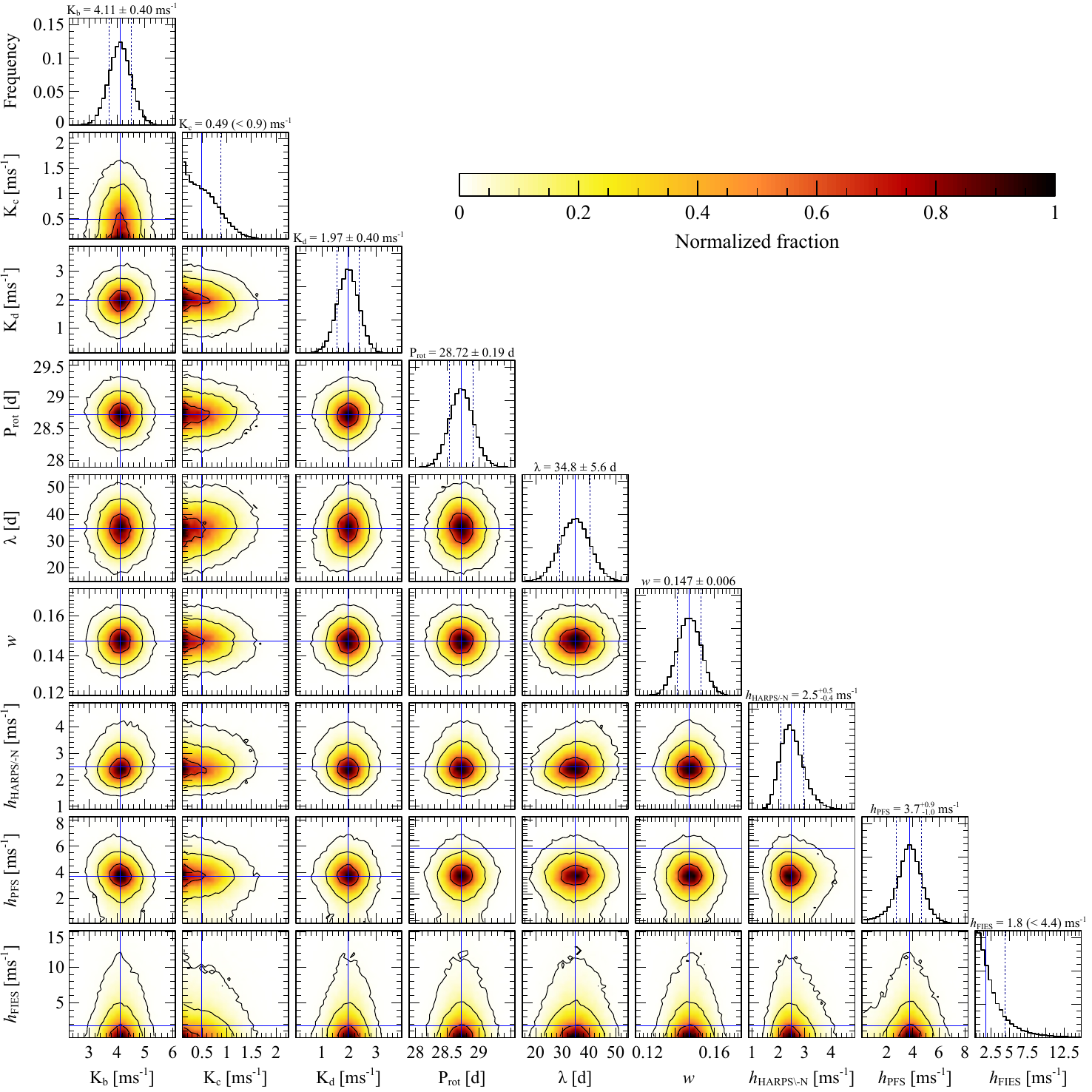}
    \caption{Posterior distributions of some of the fitted parameters determined by the analysis of the GJ 9827 RVs (Analysis 1 in Table \ref{tab:rvanalysis}). For the sake of readability, we only show the RV semi-amplitude of the planets and the GP hyper parameters.}
    \label{fig:corner}
\end{center}    
\end{figure*}

Our results are shown in Table \ref{tab:rvanalysis}. The table shows the quantities derived from the RVs (radial velocity semi-amplitude, $K$, planet mass, $M_p$, and mean density, $\rho$) and also shows the resulting stellar activity indicators, the uncorrelated jitter, and the RV offset for each dataset.  The posterior distributions of some of the fitted parameters from Analysis 1 are shown in Figure \ref{fig:corner}.  For the sake of readability, only the RV semi-amplitude of the planets and the GP hyper parameters are reported.  The confidence
intervals of the posteriors are computed by taking the 15.87$^{\rm th}$
and 84.14$^{\rm th}$ percentiles of the distribution, except for $K_{\rm c}$ and $h_{\rm FIES}$, for which we report the median and the 84.14$^{\rm th}$ percentile.

As discussed above, the two analyses were one in which the activity priors were set by the results of Section \ref{sec:stellaractivity}, and one in which we used the results of Section \ref{sec:stellaractivity} to set the region where we'd expect the activity parameters to lie, but set the priors to have a much broader range than that suggested by the results presented in Section \ref{sec:stellaractivity}. Table \ref{tab:rvanalysis}, shows that the results of these two analyses are consistent.  Since the stellar activity indicators derived from the $K2$ lightcurve, presented in Section \ref{sec:stellaractivity}, probably best represent the stellar activity, we will focus primarily on the results from Analysis 1.

\begin{table*}
  \caption[]{Best-fit solutions from the RV analyses.}
         \label{tab:rvanalysis}
   \begin{center}
   \begin{tabular}{lcc}
            \hline\hline
            \noalign{\smallskip}
            Parameter     &  \multicolumn{2}{c}{Best-fit value}  \\
                          & Analysis 1$^a$ & Analysis 2$^b$ \\
            \noalign{\smallskip}
            \hline
            \noalign{\smallskip}
            \textbf{Stellar activity GP model.} & &  \\
            (Does not include those from the & &  \\ 
            $K2$ analysis presented in Table \ref{tab:stellaractivity}.) & &  \\
            \noalign{\smallskip}
            $h_{\rm HARPS-N, HARPS}$ [m\,s$^{-1}$] & 2.49$^{+0.48}_{-0.39}$ & $2.85^{+0.66}_{-0.51}$ \\
            \noalign{\smallskip}
            $h_{\rm FIES}$ [m\,s$^{-1}$] & 1.76$^{+2.67}_{-1.21}$ & $2.09^{+3.64}_{-1.46}$ \\
            \noalign{\smallskip}
            $h_{\rm PFS}$ [m\,s$^{-1}$] & 3.73$^{+0.93}_{-1.03}$ & $3.88^{+0.95}_{-0.95}$ \\
            \noalign{\smallskip}
            $\lambda$ [days] & 34.77$^{+5.57}_{-5.64}$  & $30.97^{+12.47}_{-11.79}$ \\ 
            \noalign{\smallskip}
            $w$ & 0.147$\pm$0.006 & $0.196^{+0.041}_{-0.038}$ \\ 
            \noalign{\smallskip}
            $P_{\rm rot}$ [days] & $28.72^{+0.19}_{-0.19}$ & $30.13^{+11.11}_{-2.02}$ \\ 
            \noalign{\smallskip}
            \hline
            \noalign{\smallskip}
            \textbf{Uncorrelated jitter} & & \\
            $\sigma_{\rm jit, HARPS-N}$ [m\,s$^{-1}$] & 0.59$^{+0.40}_{-0.37}$ & $0.59^{+0.40}_{-0.38}$ \\ 
            \noalign{\smallskip}
            $\sigma_{\rm jit, HARPS}$ [m\,s$^{-1}$] & 0.80$^{+0.42}_{-0.44}$ & 0.81$^{+0.41}_{-0.44}$ \\
            \noalign{\smallskip}
            $\sigma_{\rm jit, FIES}$ [m\,s$^{-1}$] & 1.23$^{+1.58}_{-0.85}$ & $1.24^{+1.65}_{-0.86}$ \\ 
            \noalign{\smallskip}
            $\sigma_{\rm jit, PFS}$ [m\,s$^{-1}$] & 2.32$^{+1.28}_{-1.18}$ & $2.17^{+1.17}_{-1.04}$  \\ 
            \noalign{\smallskip}
            \hline
            \noalign{\smallskip}
            \textbf{RV offset} & & \\
            $\gamma_{\rm HARPS-N}$ [m\,s$^{-1}$] & 31949.335$^{+0.775}_{-0.753}$ & $31949.473^{+0.970}_{-0.916}$ \\ 
            \noalign{\smallskip}
            $\gamma_{\rm HARPS}$ [m\,s$^{-1}$] & 31948.292$^{+0.907}_{-0.876}$ & 31948.556$^{+1.103}_{-1.027}$   \\
            \noalign{\smallskip}
            $\gamma_{\rm FIES}$ [m\,s$^{-1}$] & 31775.640$^{+1.969}_{-1.988}$ & $31775.623^{+2.222}_{-2.192}$ \\ 
            \noalign{\smallskip}
            $\gamma_{\rm PFS}$ [m\,s$^{-1}$] & 0.447$^{+0.984}_{-0.979}$  & $0.533^{+1.019}_{-0.988}$  \\ 
            \noalign{\smallskip}
            \hline
            \noalign{\smallskip}
            \noalign{\smallskip}
            \textbf{Quantities derived from RVs} & & \\
            \noalign{\smallskip}
            $K_{\rm b}$ [m\,s$^{-1}$] & 4.11$^{+0.40}_{-0.40}$  & $4.10^{+0.37}_{-0.37}$  \\ 
            \noalign{\smallskip}
            $K_{\rm c}$$^1$ [m\,s$^{-1}$] & $0.49 (< 0.87)$  & $0.39 (< 0.74)$  \\ 
            \noalign{\smallskip}
            $K_{\rm d}$ [m\,s$^{-1}$] & 1.97$^{+0.40}_{-0.40}$  & $1.80^{+0.43}_{-0.48}$ \\ 
            \noalign{\smallskip}
            $M_{\rm p,b}$ ($\mearth$) & 4.91$^{+0.49}_{-0.49}$ & $4.90^{+0.45}_{-0.45}$  \\
            \noalign{\smallskip} 
            $M_{\rm p,c}$$^1$ ($\mearth$) & $0.84 (< 1.50)$ & $0.67 (< 1.27)$  \\ 
            \noalign{\smallskip} 
            $M_{\rm p,d}$ ($\mearth$) & 4.04$^{+0.82}_{-0.84}$ & $3.71^{+0.90}_{-0.99}$ \\ 
            \noalign{\smallskip}                          
            $\rho_{\rm b}$$^2$ [g\,cm$^{-3}$] & $6.93^{+0.82}_{-0.76}$ & $6.90^{+0.76}_{-0.71}$  \\ 
            \noalign{\smallskip} 
            $\rho_{\rm c}$$^{1,2}$ [g\,cm$^{-3}$] & $2.42 (< 4.35) $ & $1.93 (< 3.66) $ \\  
            \noalign{\smallskip} 
            $\rho_{\rm d}$$^2$ [g\,cm$^{-3}$] & $2.69^{+0.58}_{-0.57}$ & $2.46^{+0.63}_{-0.66}$ \\ 
            \noalign{\smallskip}
            \hline
     \end{tabular}   
     \end{center}
\begin{flushleft}
$^a$ \textrm{PyORBIT} analysis in which we set priors on $P_{\rm rot}$, $\lambda$, and $w$ from the stellar activity analysis using the $K2$ lightcurve only. See Section \ref{sec:stellaractivity} and Table \ref{tab:stellaractivity}.
$^b$ \textrm{PyORBIT} analysis in which we use the  results of the activity analysis described in \ref{sec:stellaractivity} to guide our choice of priors, rather than using these results exactly. Specifically, we impose Gaussian priors on $P_{\rm rot}$, $\lambda$, and $w$ with $P_{\rm rot} = 35 \pm 10$ days, $\lambda = 36 \pm 15$ days, and $w = 0.15 \pm 0.05$.
$^1$ For upper limits, we report the median and the 84$^{\rm th}$ percentile.
$^2$ The density was determined by sampling the posterior distributions for the mass and radius, and presenting the median, 16$^{\rm th}$ and 84$^{\rm th}$ percentiles of the resulting distribution.
\end{flushleft}
\end{table*}

\begin{figure}
\begin{center}
\includegraphics[width=9.0cm]{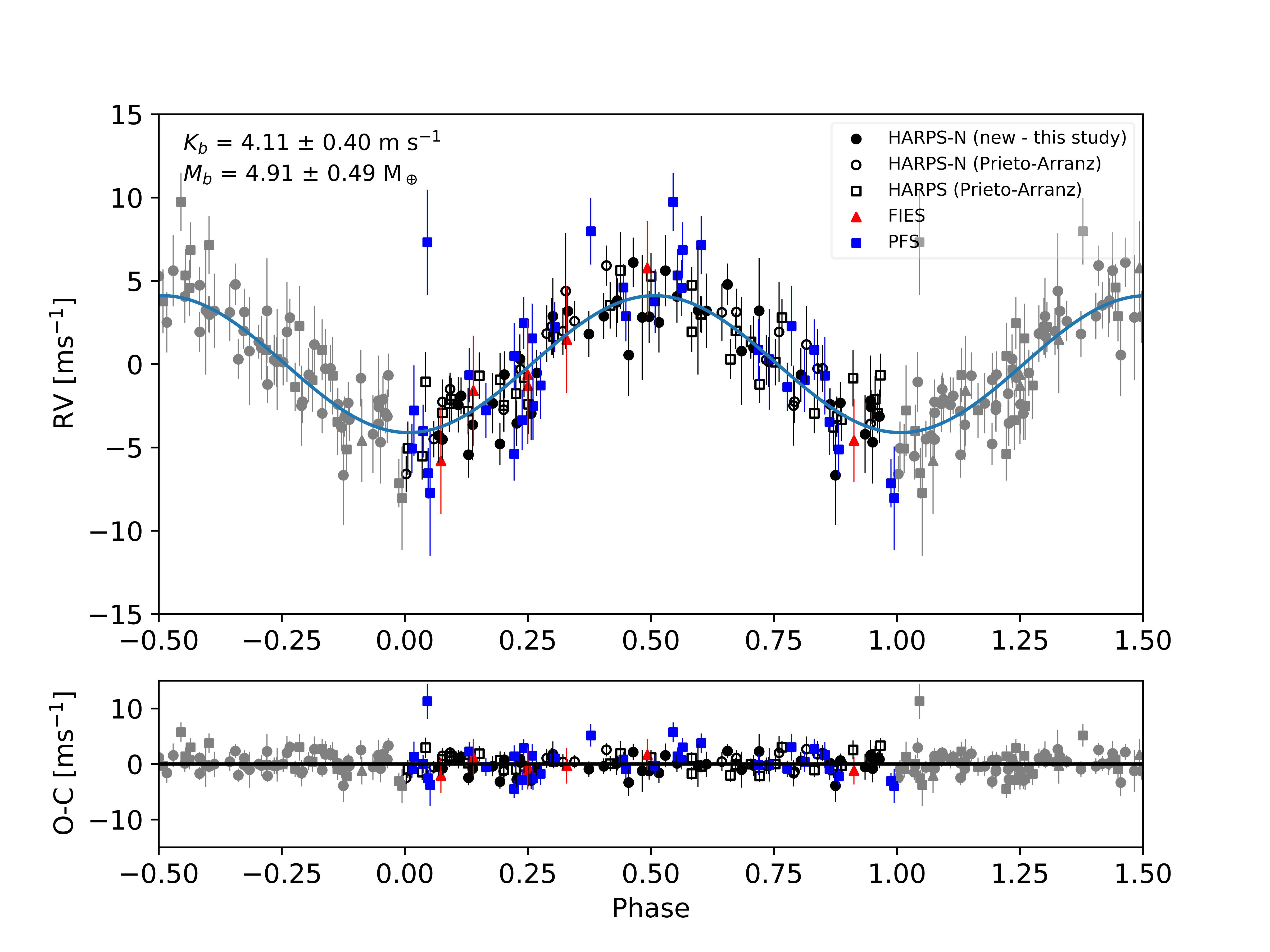}
\includegraphics[width=9.0cm]{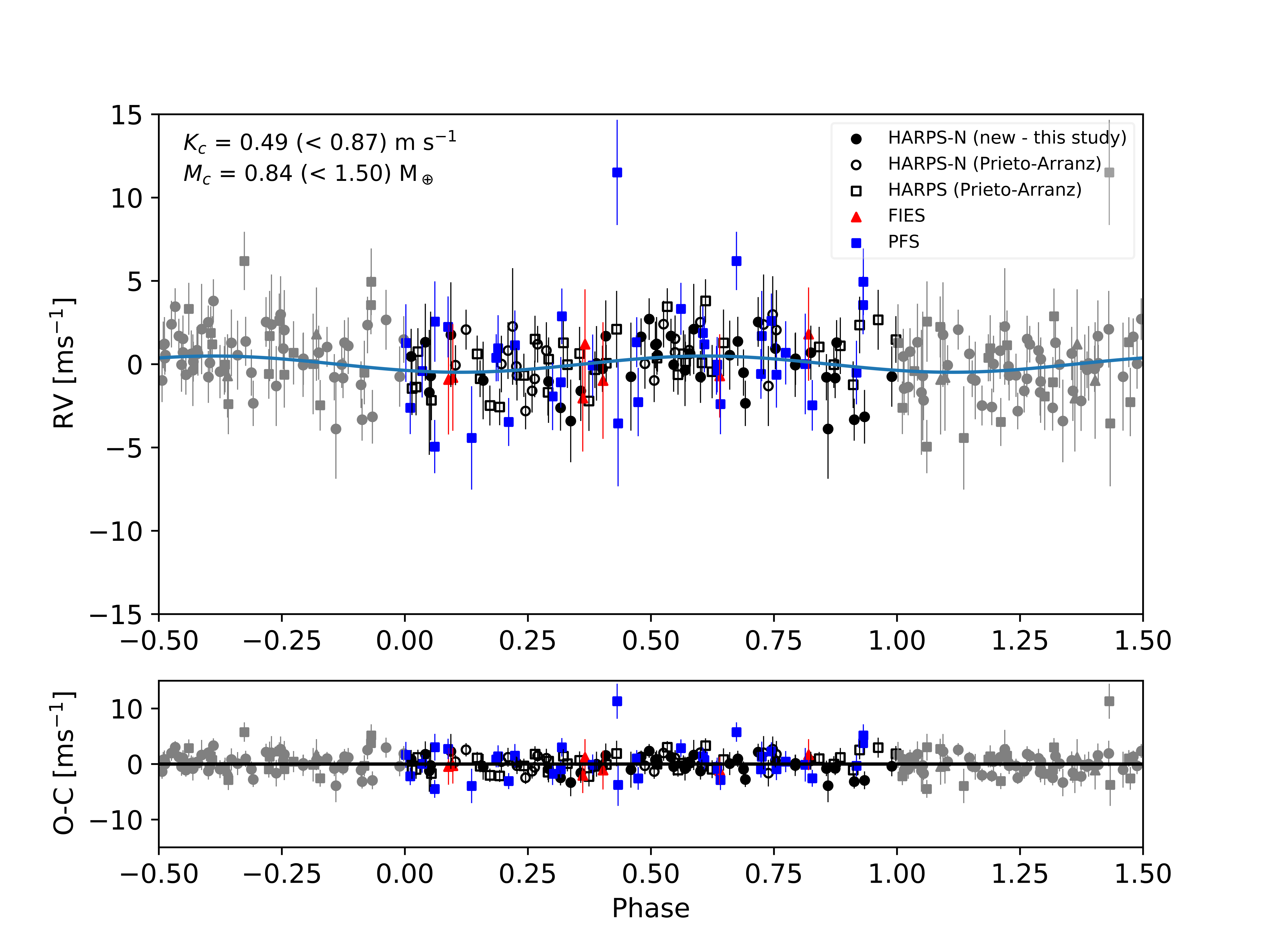}
\includegraphics[width=9.0cm]{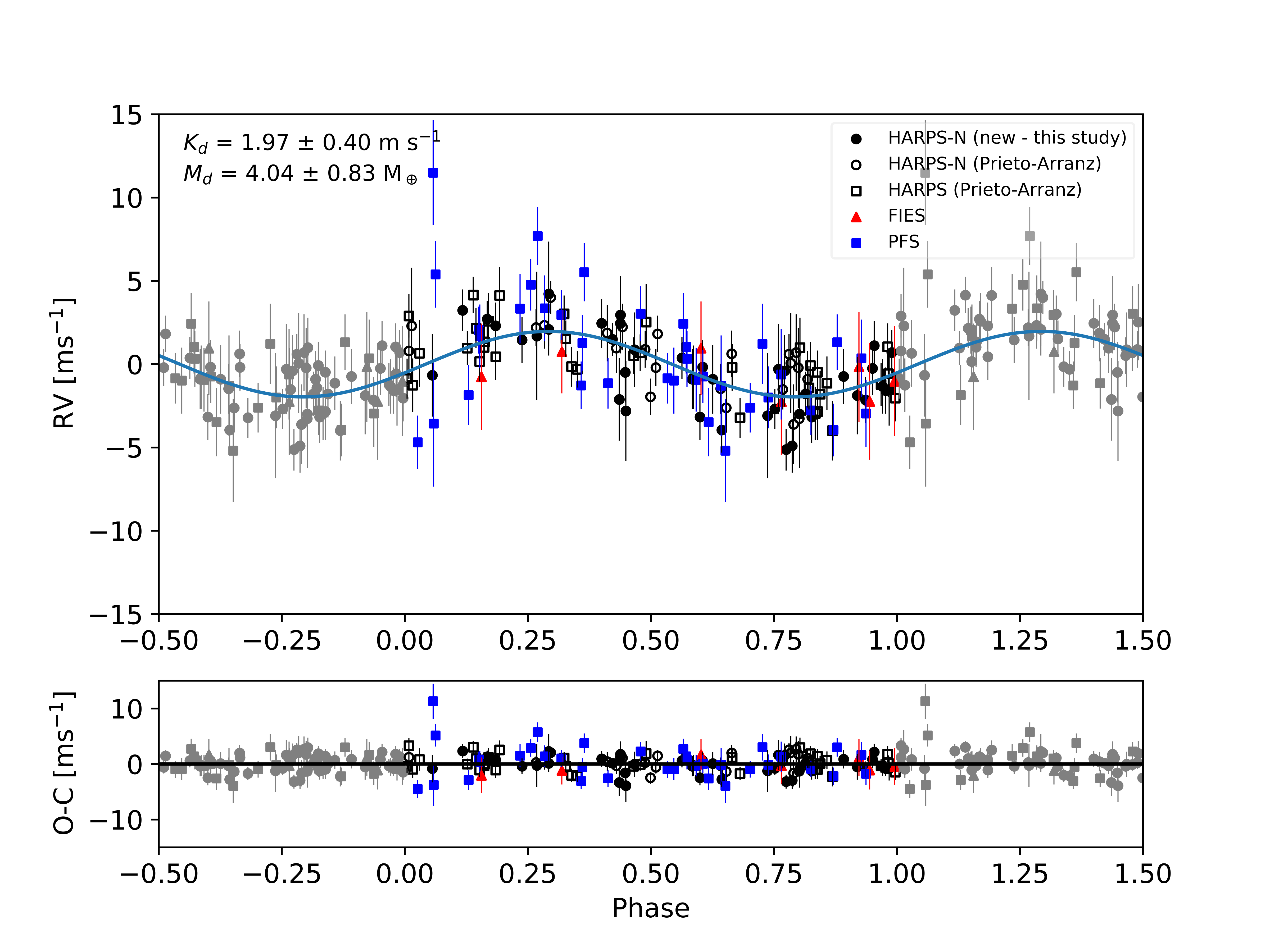}
\caption{Orbital solutions and RV residuals for GJ 9827 b (top panel), GJ 9827 c (middle panel), and GJ 9827 d (lower panel), phased on the period of the corresponding planet and with the RV contributions from the other planets removed. The details are discussed in Section \ref{sec:rvanalysis}, and these figures show the results from Analysis 1.}
\label{fig:rvHPF} 
\end{center}
\end{figure}

Figure \ref{fig:rvHPF} shows the orbital solutions and RV residuals from Analysis 1, for GJ 9827 b (top panel), GJ 9827 c (middle panel), and GJ 9827 d (lower panel), phased on the period of the corresponding planet and after removing the RV contributions from stellar activity and from the other planets. GJ 9827 b has a RV semi amplitude of $K_b = 4.11 \pm 0.40$ \ms, suggesting a mass of $M_{p,b} = 4.91 \pm 0.49$ M$_\oplus$.  The RV semi amplitude, $K_c$, for GJ 9827 c is small and suggests a mass of $M_{p,c} = 0.84$ M$_\oplus$ with an upper limit of $1.50$ M$_\oplus$, while for GJ 9827 d the RV semi amplitude is $K_d = 1.97 \pm 0.40$ \ms with a resulting mass
estimate of $M_{p,d} = 4.04^{+0.82}_{-0.84}$ M$_\oplus$. The mass estimate for GJ 9827 b therefore has a precision of better than 10\%, while that for GJ 9827 d is close to 20\%.

\begin{figure}
\begin{center}
     \includegraphics[width=9.25cm]{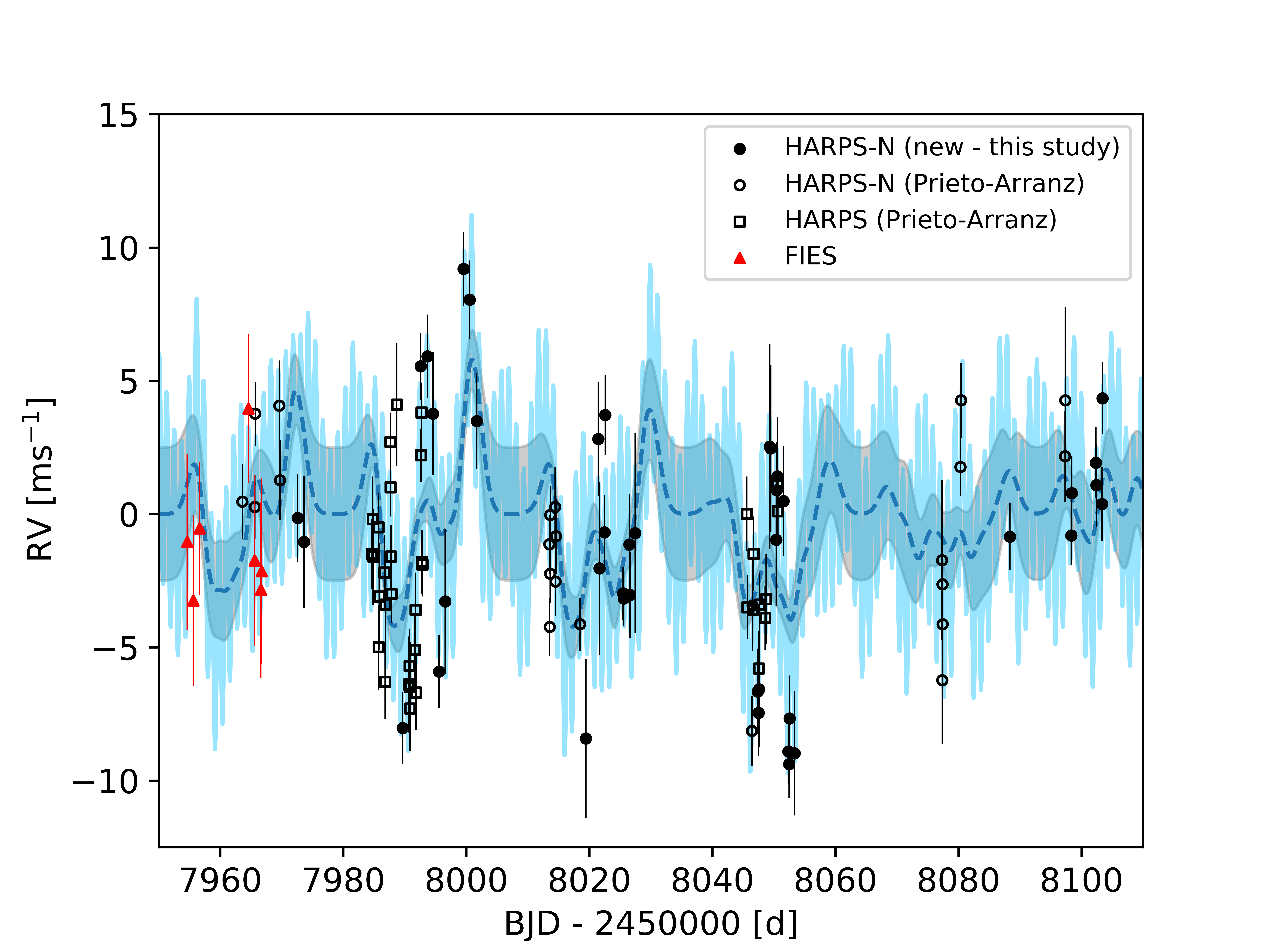}
     \caption{HARPS-N (filled and open circles), HARPS (open squares), and FIES (red triangles) RVs, together with the best-fit models which includes the planets' signals and the GP model of the correlated stellar noise (light blue curve).  Also shown is the GP solution (dashed blue curve) and its associated uncertainty range (grey shaded region).}
     \label{fig:RV_GPs}
\end{center}
\end{figure}

Figure \ref{fig:RV_GPs} shows the HARPS-N (filled and open circles), HARPS (open squares), and FIES (red triangles) RVs, together with the best-fit model which includes the planets' signals and the GP model of the correlated stellar noise (light blue curve). Also shown is the GP solution (dashed blue curve) and its associated uncertainty range (grey shaded region).  

We don't, however, show the PFS RVs in Figure \ref{fig:RV_GPs}. What Figure \ref{fig:rvHPF} shows is that the RV residuals for some of the PFS data are considerably larger than that for the other datasets.  This is most likely because the PFS data covers a long time interval and, during some periods, is insufficiently well sampled to constrain the stellar activity.

To test the consequences of this, we carried out two more analyses, both using the same activity priors as used by Analysis 1 in Table \ref{tab:rvanalysis}. In one we excluded PFS data that appeared to be insufficiently well sampled to constrain the stellar activity, and in the other we used HARPS-N and HARPS data only. In the first of these analyses, we retained the 6 PFS RVs between BJD=2455428.80 and BJD=2455439.82, the 12 PFS RVs between BJD=2455785.72 and BJD=2455485.70, and the 3 PFS RVs between BJD=2456139.86 and BJD=2456150.83.  In both cases, the results were consistent with, and of a similar precision to, those presented in Table \ref{tab:rvanalysis}.  Consequently, we conclude that the PFS sampling does not significantly influence our results, both in terms of the best estimate or the precision.

We also carried out one additional analysis, using the same activity priors as in Analysis 1, in which we relax the constraint that the planet eccentricities are all zero.  We do, however, constrain the eccentricities to be less than 0.2, which is based on pure N-body simulations using {\tt mercury6} that indicate that this is required for stability.  The results from this analysis do allow for the planets to have small eccentricities, but the resulting RV semi-amplitudes, and planet masses, are very close to those produced by the equivalent analysis with circular orbits. The resulting eccentricities are also consistent with $e = 0$ at $2.45\sigma$ which suggests that this result is not significant \citep{lucy71}. There is therefore no strong evidence to indicate that the orbits are non-circular.

\section{Discussion}
\label{sec:discussion}
Our analysis has allowed us to estimate the masses of GJ 9827 b and d with a precision of better than 10\% and close to 20\%, respectively.  We can't, however, put a strong constraint on the mass of GJ 9827 c.  Our analysis suggests an upper limit (84\%) for GJ 9827 c's RV semi-amplitude of $< 1$ \ms.  If we assume an Earth-like composition ($M_p \simeq 1.9 M_\oplus$, similar to Kepler-78b, \citealt{pepe13}), the RV semi-amplitude for this planet would be $\simeq 1$ \ms.  This might suggest that GJ9827c is unlikely to have an Earth-like composition.  

\begin{figure}
\begin{center}
\includegraphics[width=8.5cm]{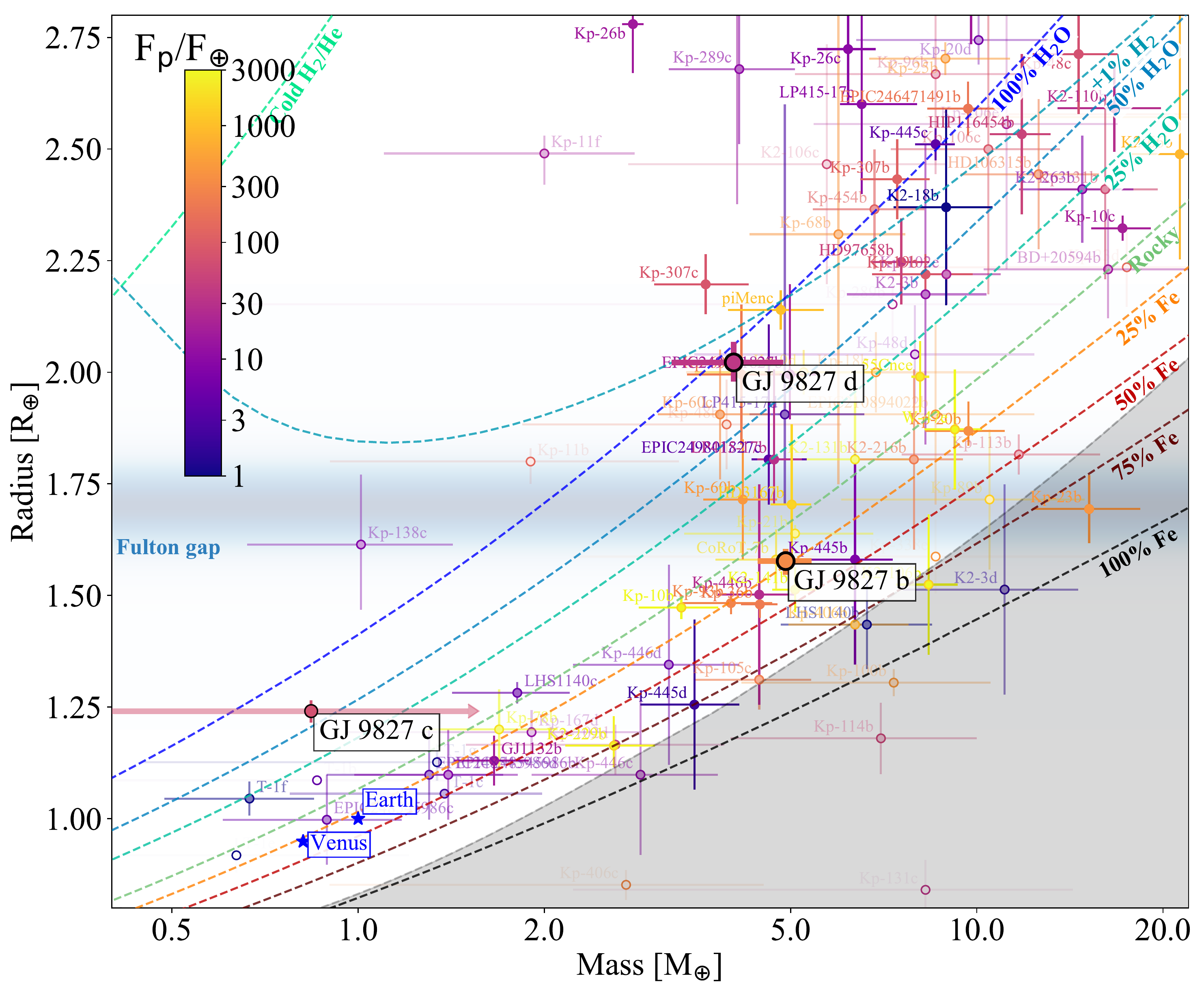}
\caption{Mass-Radius diagram for GJ 9827 b, GJ 9827 c, and GJ 9827 d together with all planets with a measured mass and radius from the Extrasolar Planets Encyclopaedia. The dashed lines show different compositions, taken from \citet{zeng16}, plus one additional as yet unpublished track for a planet in which H$_2$ makes up 1\% of its mass.  The data points are shaded according to the precision of their mass estimate and are color-coded according to their incident flux. Also shown are Earth and Venus, for reference, and we indicate the approximate location of the radius gap \citep{fulton17}.}
\label{fig:mass_radius_diagram} 
\end{center}
\end{figure}

Figure \ref{fig:mass_radius_diagram} shows GJ 9827 b, c, and d on a mass-radius diagram which also includes all planets with a measured mass and radius from the Extrasolar Planets Encyclopaedia\footnote{Available at \url{http://www.exoplanet.eu}}. The data points are shaded according to the precision of their mass estimate and are color-coded according to their incident flux, relative to that of the Earth.  The dashed lines show different compositions, taken from \citet{zeng16}, plus one as yet unpublished track for a planet in which H$_2$ makes up 1\% of its mass.  The figure also shows the Earth and Venus, for reference, and indicates the approximate location of the radius gap \citep{fulton17}.

GJ 9827 b is consistent with having a rocky, terrestrial (Earth-like) composition.  The result for GJ 9827 c suggests that it is not consistent with being rocky, and that water could still make up a substantial fraction of its mass.  There are, however, indications that non-detections, like that of GJ 9827 c, could return RV semi-amplitudes that are biased low with respect to the real RV semi-amplitudes \citep[e.g.,][]{damasso18}. Therefore, our results cannot be interpreted as strong evidence for GJ 9827 c not being rocky.  GJ 9827 b, on the other hand, would seem to be composed mostly of silicates and iron.  The best estimate suggests that its iron core makes up about 25\% of its mass, similar to that for the Earth and Venus, and the density estimate would seem to rule out GJ 9827 b having H/He on its surface, or the presence of a thick envelope of volatiles. 

The bulk density of GJ 9827 d, and its location in the mass-radius diagram (Figure \ref{fig:mass_radius_diagram}), suggests that it probably does retain a reasonably substantial atmosphere, with water potentially making up a substantial fraction of its mass.  GJ 9827 would therefore appear to host a super-Earth that is probably rocky (GJ 9827 b) and one (GJ 9827 d) that probably retains a substantial atmosphere.  These two planets appear to bracket the radius gap suggested by \citet{fulton17} and \citet{vaneylen18}.  

The stellar fluxes received by GJ 9827 b and c are about 316 and 73 times that received by the Earth, respectively.  If they are both rocky, they may still have formed with a composition similar to that of GJ 9827 d, but may have since lost their atmospheres through photo-evaporation \citep{lopezfort14}. If, however, water still makes up a substantial fraction of GJ 9827 c's mass, then this could have implications for the formation of this system.  It could suggest that GJ 9827 c and d both formed beyond the snowline, with GJ 9287 b forming inside the snowline.  Migration could then have produced the configuration we see today. That the system is in a near 1:3:5 resonance \citep{prieto-arranz18} might be consistent with this scenario.  In such a scenario GJ 9827 c could still retain a water-rich atmosphere even at its current level of irradiation \citep{lopez17}. 

On the other hand, the stellar flux received by GJ 9827 d is about 36 times that received by the Earth, which may not be sufficient for GJ 9827 d to have lost much of its primordial atmosphere \citep{owenwu13,lopez13}, whether water-rich or predominantly H/He \citep{lopez17}.  This system may therefore be consistent with photo-evaporation playing a key role in generating the radius gap suggested by \citet{owenwu13} and \citet{lopez13}, and first detected by \citet{fulton17}.  

In fact, if those planets above the radius gap typically retain a H/He atmosphere, then a prediction of the photo-evaporation model is that planets just above, and just below, the radius gap should have similar masses, since the envelope should make up only a small fraction of the mass of those just above the gap \citep{lopez18}. The similar masses of GJ 9827 b and d are intriguingly consistent with this prediction. 

There are, however, alternative explanations.  For example, the luminosity of the cooling core could completely erode light envelopes, while having little impact on heavier envelopes \citep{ginzburg18}. This would produce a deficit of intermediate-mass planets and, hence, may also explain the observed radius gap \citep{fulton17}.  Systems like GJ 9827 will therefore play a key role in determining which of these scenarios most likely explains this radius gap.    


\subsection{The composition of planets below the radius gap}
\label{sec:composition}
One of the goals of the HARPS-N Collaboration is to try to determine the typical composition of planets with radii similar to that of the Earth.  In particular, are planets below the radius gap first clearly mapped by \citet{fulton17} rocky?  It has already been suggested \citep{rogers15} that most planets above this gap still retain significant envelopes of volatiles, but it's not yet clear if most planets below the gap are primarily composed of silicates and iron.

\begin{figure}
    \centering
    \includegraphics[width=8.5cm]{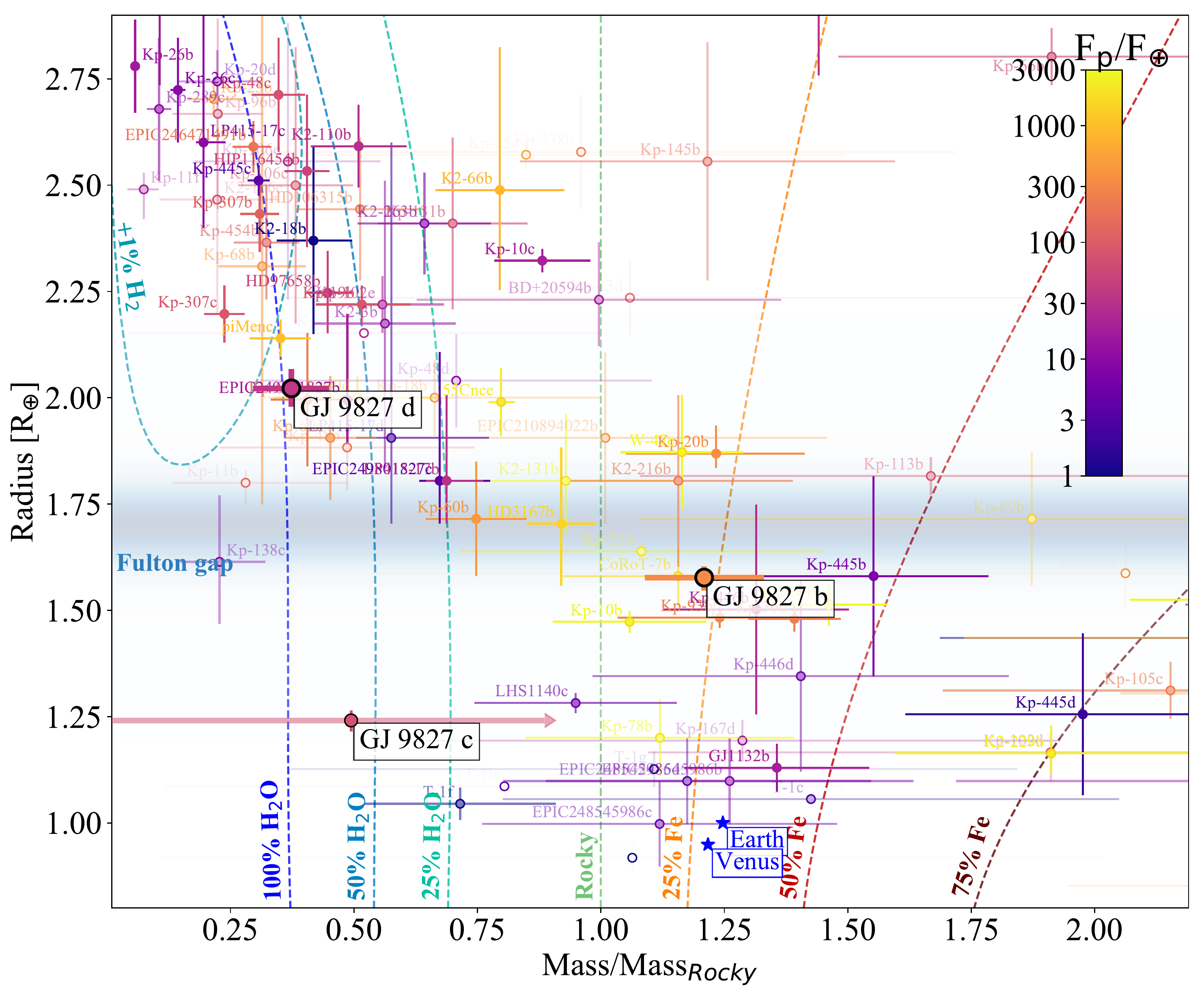}
    \caption{Similar to Figure \ref{fig:mass_radius_diagram}, except the  masses are scaled according to the minimum mass they would need, given their radius, to be rocky.  As in Figure \ref{fig:mass_radius_diagram}, the data points are shaded according to the precision of their mass estimate and color-coded according to their incident flux, relative to that of the Earth. It also illustrates the location of the radius gap \citep{fulton17}.}
    \label{fig:rescaledM}
\end{figure}

In Figure \ref{fig:rescaledM} we plot the same data as in Figure \ref{fig:mass_radius_diagram},  but scale the planet masses according to the minimum mass they would need, given their radius, in order to be rocky (see composition curves in Figure \ref{fig:mass_radius_diagram}). As in Figure \ref{fig:mass_radius_diagram}, the data points are shaded according to the precision of their mass estimate and are color-coded according to their incident flux, relative to the Earth. We also show the approximate location of the radius gap \citep{fulton17}.

What Figure \ref{fig:rescaledM} shows quite clearly is that those with radii above the gap, including GJ 9827 d, tend to have masses below that required for them to be rocky, while those below the gap tend to have masses above the mass at which they would be rocky.   Our estimate for GJ 9827 b suggests that it is clearly rocky.  The upper limit for GJ 9827 c suggests that it is not rocky and that it may still retains a reasonable amount of water, and other volatiles.  However, as highlighted in \citet{damasso18}, there are indications that a result like that for GJ 9827 c could be biased low, so we really can't rule out that GJ 9827 c is indeed rocky.

However, what Figure \ref{fig:rescaledM} also shows is that the only other known planet below the radius gap that is inconsistent with being rocky at $1 \sigma$ is Trappist-1f.  Trappist-1f is, however, around a very-low-mass star and has a low bolometric irradiation ($M_* \sim 0.08 M_\odot$ and $F_p/F_\oplus \sim 0.382$, \citealt{gillon17}). It is quite strongly irradiated in the XUV \citep{wheatley17,bolmont17}, but probably does still retain a volatile-rich envelope \citep{quarles17}. If GJ 9827 c does indeed still retain a substantial gaseous envelope, then it would be one of the most heavily irradiated planets below the radius gap to do so, and the only one orbiting an FGK star.

\section{Conclusions}
\label{sec:conclusions}
Here we present the results of our analysis of the GJ 9827 planetary-system, a system already known to contain 3 super-Earths \citep{niraula17,rodriguez18}.  We repeat the $K2$ lightcurve analysis and recover planetary radii that are consistent with these earlier analyses.  We then carry out an RV analyses using the Magellan/PFS and FIES radial velocities first presented by \citet{teske18} and \citet{niraula17} respectively, the HARPS and HARPS-N radial velocities presented by \citep{prieto-arranz18}, and with 41 additional new RV observations from HARPS-N \citep{cosentino12}. 

Although our RV analysis can't provide a strong constraint on the mass of GJ 9827 c, we can estimate the masses of GJ 9827 b and d with precisions of better than 10\% (b) and close to 20\% (d).  We find that GJ 9827 b is probably rocky, with an iron core, but is unlikely to have a mass as high as suggested by \citet{teske18}. GJ 9827 d, on the other hand, almost certainly retains a significant envelope of volatiles.  

Using HARPS, HARPS-N and FIES RVs, \citet{prieto-arranz18} also estimated the masses of the planets in the GJ 9827 system.  They conclude that GJ 9827 b is probably rocky, with an iron core, and that GJ 9827 d still retains an evelope of volatiles, which is consistent with the results presented here.  However, our estimates for the mass of GJ 9827 b and GJ 9827 d are inconsistent with their estimates at the 1$\sigma$ level.  Our analysis suggests that both GJ 9827 b and GJ 9827 d have higher masses than suggested by \citet{prieto-arranz18}.  Their estimate for GJ 9827 b is still consistent with an Earth-like composition, but their estimate for GJ 9827 d would seem to suggest a much lower density than is suggested by our analysis.

\citet{prieto-arranz18} also claim a 2$\sigma$ detection for the mass of GJ 9827 c, while we can only really set an upper limit. Although our upper limit is consistent, at 1$\sigma$, with their result, their analysis suggests that GJ 9827 c may well be rocky, whereas ours suggests that it probably is not.  It would seem quite important to understand this difference, since the composition of GJ 9827 c could constrain where the planets in this system formed.  If water makes up a significant fraction of its mass, then that might suggest that the outer planets in this system formed beyond the snowline. If not, then {\it in situ} formation is still a possibility \citep{chiang2013}.  It is possible, however, that our non-detection has returned RV semi-amplitudes that are biased low \citep{damasso18}.

GJ 9827 is particularly interesting system since it hosts a rocky super-Earth near the lower boundary of the radius gap detected by \citet{fulton17} and one that retains a substantial atmosphere near the upper boundary of this gap.  Consequently, this system could be consistent with the inner-most one being sufficiently strongly irradiated to have lost its atmosphere via photo-evaporation \citep{lopez17,owenwu17}.  If GJ 9827 d retains a low-mass H/He envelope, rather than a water-rich atmosphere, then GJ 9827 b and d having similar masses is also consistent with the photoevaporation model. However, we can't yet exclude alternative explanations, such as the luminosity of the cooling core eroding the lighter envelopes \citep{ginzburg16,ginzburg18}.  Therefore, understanding systems like GJ 9827 will help to determine which scenario is most likely.

Our results also have implications for the typical composition of planets below the radius gap detected by \citet{fulton17}.  Most planets with well-constrained masses below this radius gap have compositions consistent with them being rocky.  This is indeed the case for GJ 9827 b, but our analysis can't rule out that GJ 9827 c still retains a water-rich atmosphere. However, if this is the case, GJ 9827 c would be the one of the most heavily irradiated super-Earths below the radius gap that still retains a substantial volatile envelope.  Given the faintness of the star (V=10.3) and the expected RV amplitude ($1$ \ms \ for a rocky composition), it seems likely that only the next generation of high-precision velocimeters on large telescopes, such as ESPRESSO \citep{pepe10} or G-CLEF \citep{szentgyorgyi12}, will allow a mass determination that has sufficient precision (better than $\sim 20$\%) to uncover its internal composition.

\vspace*{0.25cm}
 As already highlighted by \citet{rodriguez18} and \citet{niraula17}, GJ 9827 is bright, and cool, and hence is a potential target for atmospheric characterisation via transit spectroscopy \citep{seager00}.  The expected signal can be calculated from the planet and star's radii, and the scale-height of the planet's atmosphere \citep{vanderburg16c}.  Our analysis suggests that if both GJ 9827 d and GJ 9827 c have predominantly H/He envelopes, the atmospheric signal could be as high as a few 100 ppm, which could be detected by the {\it Hubble Space Telescope}.  However, if their atmospheres are predominantly water, a detection may require the {\it James Webb Space Telescope}. That GJ 9827 probably hosts a rocky super-Earth and one that probably retains a substantial atmosphere, and that these two planets bracket the radius gap detected by \citep{fulton17}, makes it a particularly interesting target. 

\begin{table*}
\caption{HARPS-N radial velocity data.}
\label{tab:harpsndata}
\begin{tabular}{lllllll}
\hline
\hline
\noalign{\smallskip}
 BJD$_{\rm UTC}$ & RV & $\sigma_{\rm RV}$ & BIS$_{\rm span}$  & FWHM & S$_{\rm HK}$ & $\sigma_{\rm S_{\rm HK}}$ \\
      (d) & (m s$^{-1}$) & (m s$^{-1}$) & (m s$^{-1}$) & (km s$^{-1}$) & (dex) & (dex) \\
\noalign{\smallskip}
\hline
\noalign{\smallskip}
\noalign{\smallskip}
2457972.581025	&	31949.19	&	1.66	&	54.56	&	6.16146	&	0.760934	&	0.012149 \\
2457973.598897	&	31948.29	&	2.48	&	47.05	&	6.08261	&	0.703956	&	0.018964 \\
2457989.650249	&	31941.31	&	1.36	&	44.43	&	6.13164	&	0.690250	&	0.008613 \\
2457992.585770	&	31954.88	&	1.25	&	47.19	&	6.13195	&	0.712646	&	0.007336 \\
2457993.670628	&	31955.25	&	1.57	&	52.67	&	6.13304	&	0.734660	&	0.010788 \\
2457994.574853	&	31953.10	&	2.31	&	44.91	&	6.13740	&	0.716458	&	0.018600 \\
2457995.576615	&	31943.43	&	1.37	&	47.27	&	6.13503	&	0.738125	&	0.008665 \\
2457996.564509	&	31946.06	&	2.71	&	62.28	&	6.14124	&	0.703724	&	0.025410 \\
2457999.535982	&	31958.53	&	1.39	&	46.43	&	6.14647	&	0.744546	&	0.009172 \\
2458000.538638	&	31957.38	&	1.47	&	50.08	&	6.15661	&	0.758144	&	0.010064 \\
2458001.680612	&	31952.82	&	1.81	&	43.66	&	6.15685	&	0.754857	&	0.013624 \\
2458019.448280	&	31940.92	&	2.99	&	43.19	&	6.12897	&	0.680988	&	0.030242 \\
2458021.448466	&	31952.15	&	2.14	&	50.83	&	6.12614	&	0.684284	&	0.018574 \\
2458021.635511	&	31947.30	&	3.24	&	55.74	&	6.12959	&	0.709314	&	0.035249 \\
2458022.469453	&	31948.65	&	1.39	&	44.09	&	6.12319	&	0.674695	&	0.008879 \\
2458022.578398	&	31953.05	&	1.49	&	41.74	&	6.12904	&	0.692022	&	0.009983 \\
2458025.479998	&	31946.36	&	0.99	&	49.06	&	6.13471	&	0.700214	&	0.004776 \\
2458025.580458	&	31946.17	&	1.03	&	48.23	&	6.13419	&	0.709150	&	0.005181 \\
2458026.500839	&	31948.19	&	1.91	&	52.53	&	6.14859	&	0.722292	&	0.014968 \\
2458026.617340	&	31946.30	&	1.62	&	57.31	&	6.13966	&	0.715850	&	0.011669 \\
2458027.436449	&	31948.62	&	3.75	&	49.25	&	6.13452	&	0.761666	&	0.039726 \\
2458047.362466	&	31942.68	&	1.61	&	51.38	&	6.13719	&	0.734700	&	0.012005 \\
2458047.488963	&	31941.88	&	1.63	&	48.83	&	6.13660	&	0.718712	&	0.011596 \\
2458047.572523	&	31942.76	&	2.15	&	55.64	&	6.14504	&	0.719076	&	0.018844 \\
2458049.334881	&	31951.87	&	3.86	&	55.60	&	6.13803	&	0.680903	&	0.045143 \\
2458049.484976	&	31951.80	&	3.14	&	54.38	&	6.12018	&	0.682102	&	0.031959 \\
2458050.373485	&	31948.37	&	2.48	&	51.30	&	6.13536	&	0.642163	&	0.022632 \\
2458050.447982	&	31950.23	&	1.81	&	45.18	&	6.12398	&	0.686098	&	0.013886 \\
2458050.564769	&	31950.75	&	2.24	&	47.89	&	6.12488	&	0.636900	&	0.019858 \\
2458051.552113	&	31949.82	&	2.07	&	42.45	&	6.13584	&	0.675663	&	0.018119 \\
2458052.333890	&	31940.43	&	1.21	&	47.34	&	6.13214	&	0.665658	&	0.007353 \\
2458052.475119	&	31939.95	&	1.26	&	51.87	&	6.12258	&	0.650656	&	0.007588 \\
2458052.551769	&	31941.67	&	1.61	&	50.37	&	6.11954	&	0.680874	&	0.012084 \\
2458053.372143	&	31940.36	&	2.33	&	41.75	&	6.11748	&	0.669119	&	0.021128 \\
2458088.371319	&	31948.49	&	1.25	&	52.18	&	6.12414	&	0.693171	&	0.006900 \\
2458098.320396	&	31948.53	&	1.10	&	49.90	&	6.12255	&	0.702622	&	0.005724 \\
2458098.426520	&	31950.12	&	1.40	&	45.10	&	6.11918	&	0.712788	&	0.009317 \\
2458102.331265	&	31951.26	&	1.33	&	45.02	&	6.11895	&	0.704454	&	0.008007 \\
2458102.409463	&	31950.42	&	1.56	&	50.57	&	6.12284	&	0.696478	&	0.010865 \\
2458103.342309	&	31949.71	&	1.39	&	54.92	&	6.10729	&	0.749587	&	0.008941 \\
2458103.418424	&	31953.68	&	1.35	&	49.18	&	6.11461	&	0.737197	&	0.008863 \\
\noalign{\smallskip} 
\hline
\end{tabular}
\end{table*}

\section*{Acknowledgements}
L.M. and D.M. acknowledge support from INAF/Frontiera through the "Progetti Premiali" funding scheme of the Italian Ministry of Education, University, and Research. Some of this work has been carried out within the framework of the NCCR PlanetS, supported by the Swiss National Science Foundation. A.V. is supported by the NSF Graduate Research Fellowship, grant No. DGE 1144152. This work was performed in part under contract with the California Institute of Technology (Caltech)/Jet Propulsion Laboratory (JPL) funded by NASA through the Sagan Fellowship Program executed by the NASA Exoplanet Science Institute (A.V. and R.D.H.). A.C.C. acknowledges support from STFC consolidated grant number ST/M001296/1. D.W.L. acknowledges partial support from the Kepler mission under NASA Cooperative Agreement NNX13AB58A with the Smithsonian Astrophysical Observatory. C.A.W. acknowledges support by STFC grant ST/P000312/1. X.D. is grateful to the Society in Science-Branco Weiss Fellowship for its financial support. This material is based upon work supported by the National Aeronautics and Space Administration under grants No. NNX15AC90G and NNX17AB59G issued through the Exoplanets Research Program. This publication was made possible through the support of a grant from the John Templeton Foundation. The opinions expressed are those of the authors and do not necessarily reflect the views of the John Templeton Foundation.The HARPS-N project has been funded by the Prodex Program of the Swiss Space Office (SSO), the Harvard University Origins of Life Initiative (HUOLI), the Scottish Universities Physics Alliance (SUPA), the University of Geneva, the Smithsonian Astrophysical Observatory (SAO), and the Italian National Astrophysical Institute (INAF), the University of St Andrews, Queen's University Belfast, and the University of Edinburgh. This paper includes data collected by the \emph{Kepler}\ mission. Funding for the \emph{Kepler}\ mission is provided by the NASA Science Mission directorate. Some of the data presented in this paper were obtained from the Mikulski Archive for Space Telescopes (MAST). STScI is operated by the Association of Universities for Research in Astronomy, Inc., under NASA contract NAS5--26555. Support for MAST for non--HST data is provided by the NASA Office of Space Science via grant NNX13AC07G and by other grants and contracts. This research has made use of NASA's Astrophysics Data System and the NASA Exoplanet Archive, which is operated by the California Institute of Technology, under contract with the National Aeronautics and Space Administration under the Exoplanet Exploration Program.


\bibliographystyle{mnras} 
\bibliography{gj9827.bib} 

\bsp

\label{lastpage}

\end{document}